\documentclass[%
 reprint,
superscriptaddress,
 amsmath,amssymb,
 aps,pre,
floatfix
]{revtex4-2}

\usepackage{graphicx}
\usepackage{dcolumn}
\usepackage{bm}
\usepackage{color}
\definecolor{Green}{rgb}{0.10,0.60,0.00}
\usepackage{lipsum}

\renewcommand{\vec}[1]{{\bf #1}}
\newcommand{\J}{\boldsymbol{\mathcal{J}}}

\usepackage[normalem]{ulem}
\usepackage{upgreek}
\usepackage{tikz}
\usetikzlibrary{patterns}

\newcommand\Pa{P_a}
\newcommand\bfr {{\bf r}}
\newcommand\fw{_{ \rm {\scriptscriptstyle FW}}}

\definecolor{mycolor1}{RGB}{163,163,122}
\definecolor{mycolor2}{RGB}{191,205,165}
\definecolor{mycolor3}{RGB}{200,200,135}

\usepackage{hyperref}

\begin{document}


\title{Passive objects in confined active fluids: a localization transition}

\author{Ydan Ben Dor}
\author{Yariv Kafri}
\affiliation{Department of Physics, Technion -- Israel Institute of Technology, Haifa 32000, Israel}

\author{Mehran Kardar}
\affiliation{Department of Physics, Massachusetts Institute of Technology, Cambridge, Massachusetts 02139, USA}

\author{Julien Tailleur}
\affiliation{Universit\'{e} Paris Cit\'{e}, Laboratoire Mati\`{e}re et Syst\`{e}mes Complexes (MSC), UMR 7057 CNRS, F-75205 Paris, France}

\begin{abstract}
We study how walls confining active fluids interact with asymmetric passive objects placed in their bulk. We show that the objects experience non-conservative long-ranged forces mediated by the active bath. To leading order, these forces can be computed using a generalized image theorem. The walls repel asymmetric objects, irrespective of their microscopic properties or their  orientations. For circular cavities, we demonstrate how this may lead to the localization of asymmetric objects in the center of the cavity, something impossible for symmetric ones.
\end{abstract}


\maketitle

\section{Introduction}

Active matter describes systems comprising individual units that exert propelling forces on their environment~\cite{marchetti_hydrodynamics_2013,Bechinger2016RMP,o2022time}.  Examples  extend across scales, from molecular motors~\cite{schaller2010polar,sanchez2012spontaneous} to animals~\cite{ballerini2008interaction,calovi2014swarming}, including both biological~\cite{bi2016motility,matoz2017nonlinear} and artificial systems~\cite{geyer2019freezing,van2019interrupted}. Active systems have attracted a lot of interest recently due to their rich collective behaviors~\cite{vicsek_novel_1995,cates_motility-induced_2015,marchetti_hydrodynamics_2013} and to their non-trivial interactions with passive boundaries and objects~\cite{galajda_wall_2007,Tailleur2009EPL,Sokolov2010PNAS,di_leonardo_bacterial_2010,Solon2015NatPhys,granek2021anomalous,paul2022force}.
Unlike in equilibrium settings, asymmetric objects generically induce long-ranged currents in active fluids which, in turn, mediate long-range interactions between inclusions~\cite{baek_generic_2018,granek2020bodies}. These currents have been shown to play an important role in the context of motility-induced phase separation~\cite{cates_motility-induced_2015} where random obstacles placed in the bulk of a system suppress phase separation in $d<4$ dimensions~\cite{ro2021disorder}. Surprisingly, disordered obstacles localized on the boundaries also destroy phase separation in $d<3$ dimensions~\cite{bendor2021far}, something impossible in equilibrium.

\begin{figure}
\includegraphics[width=8.6 cm]{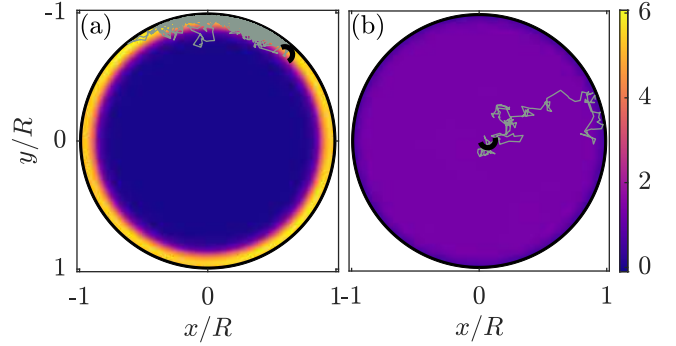}
	\caption{Probability density of a semicircular mobile object surrounded by $N=10^3$ non-interacting run-and-tumble particles confined to a circular cavity of unit diameter. Particle speed and tumbling rate are set to $v=10^{-2}$ and $\alpha=1$, respectively. The diameter of the semicircular object is equal to the particle persistence length $\ell_p=10^{-2}$. The dynamics of the object is an overdamped Brownian motion at zero temperature with translational mobility set to unity and rotational mobility set to $\gamma=10^2$ in (a) and $\gamma=10^5$ in (b). The gray lines show typical trajectories of the object. The latter is displayed in black and enlarged by a factor of six. See Appendix~\ref{app:numerics} for numerical details.}
 	\label{fig:body prob dist}
\end{figure}

In this article, we study another surprising role of boundaries. In Figure~\ref{fig:body prob dist}, we show numerical simulations of an active fluid confined in a circular cavity in which a mobile asymmetric object has been inserted. Depending on the parameters, the object is either localized close to the cavity walls or in the middle of the cavity. As we show below, this is a direct consequence of the ratchet current induced by the object in the active bath and its interactions with the cavity walls. Note that, on general symmetry grounds, an isotropic object cannot be localized in a diffusive fluid. Indeed, the sole symmetry breaking field in the vicinity of an isotropic object is the gradient of the fluid density, $\nabla \rho$. The force ${\bf F}$ exerted on the object thus satisfies ${\bf F} \propto \nabla \rho$. If the fluid is diffusive, it satisfies $\nabla^2 \rho = 0$ in the steady state so that $\nabla \cdot {\bf F}=0$~\cite{rohwer2020activated}. In analogy to Earnshaw's theorem in electrostatics, this rules out the possibility of a stable equilibrium for the passive tracer. In contrast, an asymmetric polar object introduces a symmetry-breaking vector along which it generically generates a ratchet current~\cite{Galajda2007,Tailleur2009EPL,di_leonardo_bacterial_2010,Sokolov2010PNAS,baek_generic_2018}. This current is directly related to the non-vanishing mean force ${\bf p}$ exerted by the object on the surrounding fluid~\cite{nikola_active_2016}. Due to Newton's third law, one thus generically expects a contribution to $\mathbf{F}$ along $-\mathbf{p}$, which opens up the possibility of a localization transition. Figure~\ref{fig:body prob dist} shows that this is indeed the case.

To uncover the mechanism behind this localization transition, we study the influence of boundaries on the coupling between asymmetric objects and active fluids. We start in Section~\ref{subsec:density and current} by considering the case of an asymmetric polar object in the presence of a flat confining boundary. We show that the latter alters the far-field current and density modulation induced by the polar object, and that this effect can be rationalized using a generalized image theorem. As we show in Section~\ref{subsec:force}, this leads to a repulsive force, which decays as a power-law, between the object and the wall. In Section~\ref{sec:circular cavity} we then generalize our approach to the case of a polar object confined by a circular cavity. Finally, in Section~\ref{sec:dynamics}, we consider a mobile object and show the existence of a localization transition. We note that our results could be tested experimentally by adapting a recent setup in which a \textit{symmetric} object was immersed in a circular cavity confining active colloids~\cite{paul2022force}. In this case, as expected on symmetry ground, no localization transition was observed and the interaction between the object and the wall is short ranged. We predict that employing a polar object should lead to rich physics. All derivations below are presented in two space dimensions but can easily be generalized to higher dimensions.

\section{An asymmetric object next to a flat wall}\label{sec:flat wall derivation}

\begin{center}
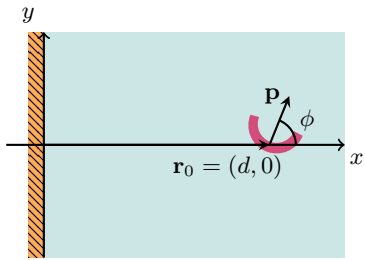
\begin{figure}[t]
    \begin{tikzpicture}
        \fill[teal!20!white] (0,-1.5) rectangle (4,1.5);
        \fill[orange!65!white] (-0.2,-1.5) rectangle (0,1.5);
        \fill[pattern=north west lines, pattern color=black] (-0.2,-1.5) rectangle (0,1.5);
        
        \fill[purple!70!white] (2.59611+0.15,0.00139995+0.4) arc(158:338:0.375) -- (3.17561+0.15,-0.217276+0.4) arc(338:158:0.25) -- cycle;
        
        \draw[-stealth,thick] (0,0) -- (3,0) node[anchor=east, yshift=-3mm, xshift=3mm] {${\bf r}_0=(d,0)$};
        \draw[-stealth,thick] (3,0) -- (3+0.26222,0.649029) node[anchor=east] {${\bf p}$};
        \draw[thick] (3.35 ,0) arc(0:68:0.35) node[anchor=west, xshift=1.5mm] {$\phi\ $};
        
        \draw[thick,->] (-.5,0) -- (4,0) node[anchor=north west] {$\!x$};
        \draw[thick,->] (0,-1.5) -- (0,1.5) node[anchor=south east] {$y$};
    \end{tikzpicture}
    \caption{An asymmetric passive object in an active fluid next to a flat wall at $x=0$. The object is located at $\vec{r}_0=(d,0)$. Due to its asymmetric shape, it experiences a force $-\vec{p}$ from the active bath and thus exerts the opposite force $\vec{p}$ on the active medium, whose orientation we denote by $\phi$.}\label{fig:configuration}
\end{figure}
\end{center}

We start by studying the influence of an infinite flat wall at $x = 0$ on an asymmetric object embedded inside the system in the neighborhood of ${\bf r}_0{= (d,0)}$ (see Fig.~\ref{fig:configuration}). We first determine in Sec. \ref{subsec:density and current} how the presence of the wall influences the ratchet current and the density modulation induced by the asymmetric object in the active bath. Then in Section~\ref{subsec:force}, we show how the density modulation translates into a net nonconservative force exerted on the object. We characterize the force in the far field limit and show its magnitude to depend on the distance from the wall and on the orientation of the object. 

In most of what follows, we focus on a dilute active bath. We thus solve for a single active particle that interacts with the obstacle and the boundaries. The average density for a bath comprising $N$ active particles is then simply $\rho_N(\bfr)=N\rho(\bfr)$ where $\rho(\bfr)$ is the probability density of finding the active particle at position $\bfr$. For the flat-wall case discussed in this section, our results are generalized to particles interacting via pairwise forces in Appendix~\ref{app:PFAPs}.

To proceed, consider the master equation for the probability density $P_a(\vec{r},\theta)$ to find an Active Brownian Particle (ABP) or a Run-and-Tumble particle (RTP) at $\vec{r}=(x,y)$ with orientation $\vec{u}(\theta)=(\cos\theta,\sin\theta)$:
\begin{align}\label{eq:FP}
    \partial_t P_a({\bf r},\theta)= & -\!\nabla\!\cdot\!\left[-\mu P_a \nabla U + v{\bf u}P_a \!-\!{D}_t\nabla P_a\right] \\
	&+{D}_r\partial_{\theta}^{2}P_a -\alpha P_a +\frac{\alpha}{2\pi} \intop d\theta' \, P_a(\vec{r},\theta') \;.\nonumber
\end{align}
Here $v$ is the self-propulsion speed of the active particle, $\mu$ its mobility, and ${D}_t$  a translational diffusivity. The particle undergoes random reorientations with a (tumbling) rate $\alpha$ and rotational diffusion with an angular diffusivity $D_r$. The object is described by the external potential $U(\vec r)$. In what follows we denote by $\tau=1/(D_r+\alpha)$ and $\ell_p\equiv v\tau$ the particle's persistence time and length, respectively. The hard wall at $x=0$ imposes a zero-flux condition: \begin{equation}\label{eq:BCJ}
    -\mu\Pa \partial_x U  + v \cos\theta \Pa - D_t\partial_x \Pa=0\;.
\end{equation}

Integrating Eq. \eqref{eq:FP} over $\theta$ leads to a conservation equation for the density field $\rho(\bfr)=\int d\theta \,P_a(\vec{r},\theta)$:
\begin{subequations}\label{eq:rho}
\begin{align}
    \partial_t \rho &= -\nabla\cdot\vec{J}\;, \\
    \vec{J} &= -\mu \rho \nabla U + v\vec{m} -{D}_t \nabla \rho\ ,\label{eq:current}
\end{align}
\end{subequations}
where $\vec{m} = \int d\theta\,\vec{u}(\theta)P_a(\vec{r},\theta)$ is the polarization of the active particle and $\vec{J}$ is the particle current in position space. The boundary condition~\eqref{eq:BCJ} then translates into $J_x(x=0,y)=0$.

The dynamics of $\vec{m}$ is then obtained by multiplying Eq.~\eqref{eq:FP} by $\vec{u}(\theta)$ and integrating over $\theta$, which gives:
\begin{equation}\label{eq:m}
        \tau\partial_t \vec{m} = \frac{\mu}{v} \nabla \cdot \boldsymbol{\sigma^a} -  \vec{m}\;, 
\end{equation}
where we have introduced the active stress tensor $\boldsymbol{\sigma^a}$~\cite{takatori_swim_2014,yang_aggregation_2014,Solon2015NatPhys,solon_pressure_2015-3,fily_mechanical_2017}:
\begin{subequations}\label{eq:sigmas}
\begin{align}
    \sigma^a_{ij} &= - \frac{v^2\tau}{2\mu}\rho \delta_{ij} + \Sigma_{ij}\;, \label{eq:sigma}\\
    \Sigma_{ij} &= -\frac{v\tau }{\mu}\left[v Q_{ij}-\left(\mu \partial_j U + D_t\partial_j\right) m_i\right]\;.\label{eq:Sigma}
\end{align}
\end{subequations}
Here $Q_{ij} = \int d\theta\,\left(u_iu_j - \delta_{ij}/2\right)P_a(\vec{r},\theta)$ is the nematic tensor and we have singled out the contribution of the ideal gas pressure $v^2\tau \rho/(2\mu)$ in the active stress tensor. 

To determine the steady-state density profile, we first note that, on large length scales and long times, far from both the confining wall and the asymmetric object, the motion of the active particle is diffusive with a diffusion coefficient ${D}_{\rm eff}={D}_{t}+v^2\tau/2$. The corresponding probability current is then given by $\vec{J} \simeq - {D}_{\rm eff} \nabla\rho$. As one moves closer to the object or the wall, this behavior is modified, which motivates us to define a residual field: the deviation $\J$ from a diffusive current~\cite{bendor2021far}
\begin{equation}\label{eq:def diff J}
    \J \equiv \vec{J} + D_{\rm eff} \nabla\rho\;.
\end{equation}
Using Eqs.~\eqref{eq:rho}-\eqref{eq:m} in the steady state where $\nabla\cdot\vec{J}=0$, one finds that the density $\rho$ satisfies
\begin{subequations}\label{eq:Poisson}
\begin{align}
    D_{\rm eff}\nabla^2\rho &= \nabla\cdot\J\;, \\
    \mathcal{J}_i &= -\mu \rho \partial_i U + \mu\partial_j\Sigma_{ij}\;. \label{eq:Poisson equation diff J}
\end{align}
\end{subequations}
The zero-flux boundary condition on the current at $x=0$ then reads
\begin{equation}\label{eq:BC}
    J_{x}(x=0,y) = \left(\mathcal{J}_{x}-D_{\rm eff}\partial_x \rho\right)\big|_{x=0} = 0\;.
\end{equation}

\subsection{Density profile and current}\label{subsec:density and current}

In the absence of the obstacle, the solution $\rho\fw(\bfr)$ to Eq.~\eqref{eq:Poisson} with the boundary condition~\eqref{eq:BC} is a homogeneous bulk complemented by a finite-size boundary layer near the wall where active particles accumulate on a scale comparable to the persistence length $\ell_p$~\cite{elgeti2009self}. In what follows, we denote by $\J\fw$ the corresponding source term in Eq.~\eqref{eq:Poisson equation diff J} and by $\boldsymbol\Sigma\fw$ the contribution~\eqref{eq:Sigma} to the active stress.
By itself, determining $\rho\fw$ is already a difficult problem, whose exact solution is not known~\cite{elgeti_wall_2013,lee2013active,Fily2015SM,ezhilan2015distribution,wagner_steady-state_2017,wagner2022steady}. To proceed, we thus work in the far field limit away from both the wall and the object, which is itself assumed to be far from the wall.

We first decompose the density field as $\rho(\bfr)=\rho\fw(\bfr)+\delta\rho(\bfr)$~\footnote{In the semi-infinite system we consider here, $\delta\rho(\bfr)$ vanishes at infinity. In a finite system, one would need to consider $\rho=z \rho\fw +\delta\rho$ to ensure the proper normalization of the density field.}. Thanks to the linearity of Poisson's equation~\eqref{eq:Poisson}, $\delta\rho$ satisfies
\begin{subequations}\label{eq:Poisson difference}
\begin{align}
    &D_{\rm eff}\nabla^2\delta\rho = \nabla\cdot\boldsymbol{\delta}\J\;, \\
    &\delta\mathcal{J}_i = -\mu \rho \partial_i U + \partial_j \delta\Sigma_{ij}\;,
\end{align}
\end{subequations}
where we have defined $\boldsymbol{\delta}\J \equiv \J-\J\fw$, and $\delta\boldsymbol{\Sigma} \equiv \boldsymbol{\Sigma}- \boldsymbol{\Sigma}\fw$. The corresponding boundary conditions read:
\begin{equation}\label{eq:BC difference}
    D_{\rm eff}\partial_x \delta\rho\big|_{x=0} = \delta\mathcal{J}_{x}(0,y)\;.
\end{equation} 
Equations~\eqref{eq:Poisson difference}-\eqref{eq:BC difference} describe the density modulation created by the asymmetric object on the density profile induced by a flat wall. 
To solve for $\delta\rho(\bfr)$, we introduce the Neumann-Green's function in the right half-plane:
\begin{equation}\label{eq:Neumann-Green's function flat wall}
    G_N(\vec{r};\vec{r}') = -\frac{1}{2\pi} \left[\ln\frac{|\vec{r}-\vec{r}'|}{\ell_p} + \ln\frac{|\vec{r}^\perp-\vec{r}'|}{\ell_p}\right]\;.
\end{equation}
Here the term involving $\vec{r}^\perp = (-x,y)$ can be interpreted as a mirror image created on the other side of the wall.
Note that the Neumann-Green's function~\eqref{eq:Neumann-Green's function flat wall} does not satisfy the boundary condition specified by Eq.~\eqref{eq:BC difference}, since its $x$--derivative vanishes on the boundary. Using Green's second identity~\cite{kevorkian1990partial,jackson_classical_1999}, this means that the solution $\delta\rho$ also includes a surface integral to enforce the correct boundary condition. All in all, it reads
\begin{subequations}\label{eq:formal solution}
\begin{align}
    \delta \rho(x,y) =& -\frac{1}{D_{\rm eff}}\intop_0^\infty dx'\intop_{-\infty}^\infty dy'\,G_N(x,y;x',y')\nabla'\cdot\boldsymbol{\delta}\J'\nonumber\\
    & - \intop_{-\infty}^\infty dy'\,G_N(x,y;0,y')\partial_x'\delta\rho'\bigg|_{x'=0} \nonumber \\
    = &  -\frac{\mu}{D_{\rm eff}}\intop_0^\infty dx'\intop_{-\infty}^\infty dy'\,\rho'\nabla'U \cdot \nabla' G_N(x,y;x',y')\label{eq:rho2}\\
    & -\frac{\mu}{D_{\rm eff}}\intop_0^\infty dx'\intop_{-\infty}^\infty dy'\, G_N(x,y;x',y') \partial_i'\partial_j'\delta\Sigma_{ij}'\label{eq:rho3} \\
    & - \frac{\mu}{D_{\rm eff}}\intop_{-\infty}^\infty dy'\,G_N(x,y;0,y')\partial_j'\delta\Sigma_{xj}'\bigg|_{x'=0}\;, \label{eq:rho4}
\end{align}
\end{subequations}
where primed derivatives are taken with respect to $x'$ and $y'$. To obtain Eq.~\eqref{eq:formal solution} we use Eqs. \eqref{eq:Poisson difference} and \eqref{eq:BC difference} and an integration by parts. As we now show, the leading-order contribution to $\delta\rho$ in the far field is given by~\eqref{eq:rho2}. Noting that $\nabla' U $ is localized at $\bfr_0=(d,0)$, we  approximate the Green's function as $\nabla' G_N(x,y;x',y')\simeq \nabla' G_N(x,y;x',y')|_{x'=d,y'=0}$ in the first integral. In the far field, where ${|\vec{r}-\vec{r}_0|,d\gg a,\ell_p}$ with $a$ the size of the object, this leads to
\begin{align}\label{eq:density force monopole flat wall}
    \rho(\vec{r}) \simeq& \rho_b + \frac{\mu }{2\pi D_{\rm eff}}\left[\frac{\vec{p}\cdot (\vec{r}-\vec{r}_0)}{\left|\vec{r}-\vec{r}_0\right|^2} + \frac{\vec{p}^\perp \cdot (\vec{r}-\vec{r}_0^\perp)}{\left|\vec{r}-\vec{r}_0^\perp
    \right|^2}\right] \nonumber \\
    & + \mathcal{O}\left(\left|\vec{r}-\vec{r}_0\right|^{-2},d^{-2}\right)\;,
\end{align}
where $\vec{p}^\perp=(-p_x,p_y)$ and we have used both that 
$\rho\fw\simeq \rho_b$ far from the wall and that ${\bf u} \cdot {\bf v} = {\bf u}^\perp \cdot {\bf v}^\perp$. In Eq.~\eqref{eq:density force monopole flat wall},  $\vec{p}$ is a force monopole defined by
\begin{equation}\label{eq:force monopole p flat wall}
    \vec{p} = -\intop d\vec{r}\,\rho\nabla U\;.
\end{equation}
It measures the force exerted on the active fluid by the object in a system without a wall, whose exact value depends on microscopic details of $U$.
Going back to Eq.~\eqref{eq:formal solution}, we show in appendix~\ref{app:multipole} that~\eqref{eq:rho3} and~\eqref{eq:rho4} are indeed negligible compared to~\eqref{eq:density force monopole flat wall}. Intuitively, this relies both on the extra derivatives in Eq.~\eqref{eq:rho3} and on the fact that we can use self-consistently the far-field approximation to $\delta \boldsymbol\Sigma$ far away from the object.

\begin{center}
\begin{figure}[t]
    \begin{tikzpicture}
        \fill[teal!20!white] (-3,-1.5) rectangle (3,1.5);
        
        \fill[purple!70!white] (2.59611-1.15,0.00139995+0.4) arc(158:338:0.375) -- (3.17561-1.15,-0.217276+0.4) arc(338:158:0.25) -- cycle;
        
        \fill[purple!70!white] (-2.59611+1.15,0.00139995+0.4) arc(22:-158:0.375) -- (-3.17561+1.15,-0.217276+0.4) arc(-158:22:0.25) -- cycle;
        
        \draw[-stealth,thick] (1.7,0) -- (1.7+0.26222,0.649029) node[anchor=east] {${\bf p}$};
        \draw[thick] (2.05 ,0) arc(0:68:0.35) node[anchor=west, xshift=1.5mm] {$\phi\ $};
        
        \draw[-stealth,thick] (-1.7,0) -- (-1.7-0.26222,0.649029) node[anchor=west, xshift=1mm] {${\bf p}^\perp$};
        \draw[thick] (-2.05 ,0) arc(180:180-68:0.35) node[anchor=east, xshift=-1mm] {$\phi\ $};
        
        \draw[thick,->] (-3,0) -- (3,0) node[anchor=north west] {$\!x$};
        \draw[thick,->] (0,-1.5) -- (0,1.5) node[anchor=south east] {$y$};
    \end{tikzpicture}
    \caption{The asymmetric object and the flat wall shown in Fig.~\ref{fig:configuration} generate  density modulations and currents in the active medium, far away from both the object and the wall, equivalent to those generated by two force monopoles $\bf p$ and $\bf p^\perp$ placed symmetrically with respect to the $x=0$ plane.}\label{fig:configuration image}
\end{figure}
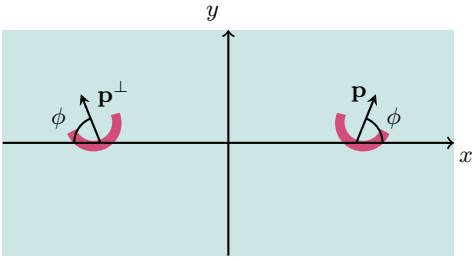
\end{center}

The far-field currents can then be obtained from the above result. We first note that, outside the object, $\J$ is negligible compared to the diffusive current $-D_{\rm eff} \nabla \rho$ (See Eq.~\eqref{eq:def diff J}). One thus has that $\vec{J}=\J-D_{\rm eff} \nabla \rho \simeq -D_{\rm eff} \nabla \rho$ so that, to leading order:
\begin{align}
    \vec{J} &\underset{|\vec{r}-\vec{r}_0|,d\gg a,\ell_p}{\simeq} \frac{\mu}{2\pi |\vec{r}-\vec{r}_0|^2}\left[\frac{2[(\vec{r}-\vec{r}_0)\cdot\vec{p}](\vec{r}-\vec{r}_0)}{|\vec{r}-\vec{r}_0|^2} - \vec{p}\right]\nonumber \\
    & + \frac{\mu }{2\pi |\vec{r}^\perp
    -\vec{r}_0|^2}\left[\frac{2[(\vec{r}-\vec{r}_0^\perp)\cdot\vec{p}^\perp
    ](\vec{r}-\vec{r}_0^\perp)}{|\vec{r}^\perp-\vec{r}_0|^2} - \vec{p}^\perp
    \right]\;.
\end{align}

In summary, to this order of the multipole expansion, the problem of finding the steady-state density in the far field is reduced to a much simpler problem
\begin{subequations}\label{eq:reduced Poisson's equation}
\begin{align}
    D_{\rm eff}\nabla^2  \rho &= \mu \nabla\cdot\left[\vec{p}\delta(\vec{r}-\vec{r}_0)\right]\;,\\
    \partial_x \rho\big|_{x=0} &= 0\;,
\end{align}
\end{subequations}
which amounts to Eqs.~\eqref{eq:Poisson difference} and~\eqref{eq:BC difference} with $\J\simeq\mu \vec{p}\delta(\vec{r}-\vec{r}_0)$. In the far field of both the object and the wall, the object thus appears as a force monopole ${\bf p}$ at position $\bfr_0$ driving the fluid while the wall is equivalent to an image monopole ${\bf p}^\perp$ at position $\bfr_0^\perp$, as can be read directly in Eq.~\eqref{eq:density force monopole flat wall} (see Fig.~\ref{fig:configuration image}).

\subsection{Nonconservative force induced on the object}\label{subsec:force}

According to Newton's third law, the object experiences a force $-{\bf p}$ from the active fluid. Equation~\eqref{eq:force monopole p flat wall} shows ${\bf p}$ to depend on the local density of active particles $\rho(\bfr)$, which in turn depends on the distance $d$ from the wall through Eq.~\eqref{eq:density force monopole flat wall}. It is thus convenient to decompose ${\bf p}$ as: 
\begin{equation}\label{eq:F definition}
    \vec{p} \equiv \vec{p}_b - \vec{F}\;,
\end{equation}
with $\vec{p}_b$ defined as the value of ${\bf p}$ when $d \to \infty$. Then $\vec{F}$ measures the change in the force due to the presence of the wall. Namely, $\vec{F}$ is the force induced on the object by the wall, which is mediated by the active bath. 

Since the interaction with the wall is equivalent, to leading order, to the interaction with an image object, we can use the results of Refs.~\cite{baek_generic_2018,granek2020bodies} to derive $\vec{F}$. In the setting considered there, two objects, referred to as object 1 and object 2, are placed at positions $\vec{r_1}$ and $\vec r_2$, with  $\vec{r}_{12}\equiv\vec{r}_1-\vec{r}_2$. When $|\vec{r}_{12}| \to \infty$ the objects experience forces $-\vec{p}_1$ and $-\vec{p}_2$ from the fluid, respectively. When $|\vec{r}_{12}|$ is finite the force experienced by object 1 is $-\vec{p}_1+\vec{F}_{12}$, with $\vec{F}_{12}$ the force exerted on object 1 due to the presence of object 2. In~\cite{baek_generic_2018}, it was shown that, to leading order in the far field, the interaction force arises due to a density modulation $\Delta \rho (\vec{r}_1)$ near object 1 due to object 2. This non-reciprocal interaction force takes the form:
\begin{equation} \label{eq:F12}
    \vec{F}_{12} = -\frac{\Delta\rho(\vec{r}_1)}{\rho_b}\vec{p}_1\;, 
\end{equation}
with
\begin{equation} \label{eq:Deltarho1}
     \Delta \rho(\vec{r}_1) = \frac{\mu}{2\pi D_{\rm eff}} \frac{\vec{r}_{12}\cdot\vec{p}_2}{|\vec{r}_{12}|^2} + \mathcal{O}(|\vec{r}_{12}|^{-2})\;.
\end{equation}
Here, $\bfr_{12}=(2d,0)$ and ${\bf p}_2={\bf p}_b^\perp$ (see Fig.~\ref{fig:configuration image}), leading to
\begin{equation} \label{eq:F12-bare}
    \vec{F} = \frac{\mu} {2 \pi D_{\rm eff} \rho_{b}} \frac{p_{b,x} }{2 d}  \vec{p}_b+\mathcal{O}(d^{-2})\;.
\end{equation}
Denoting by $\phi$ the orientation of ${\bf p}_b$ relative to the $\vec x$ axis then leads to:
\begin{align}\label{eq:force flat wall}
    \vec{F} 
    = &\frac{\mu p_b^2}{8\pi D_{\rm eff} \rho_b d} \binom{1+\cos(2\phi)}{\sin(2\phi)}+{\cal O}
    (d^{-2})\;,
\end{align} 
with $p_b = |\vec{p}_b|$. Note that this result implies that the wall \textit{always repels the object}, irrespective of its orientation $\phi$. It is easy to check that $\partial_x F_y-\partial_y F_x\neq 0$, except when $\phi\in\{0,\pi\}$, so that the interaction force is not conservative~\footnote{The divergence of the force is negative throughout the domain, $\nabla\cdot{\bf F}<0$, in principle allowing for possible stable fixed points if ${\bf F}=0$.}.

\begin{center}
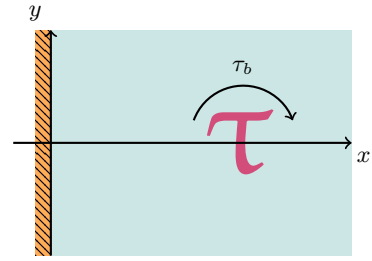
\begin{figure}[t]
    \begin{tikzpicture}[]
    
        \fill[teal!20!white] (0,-1.5) rectangle (4,1.5);
        \fill[orange!65!white] (-0.2,-1.5) rectangle (0,1.5);
        \fill[pattern=north west lines, pattern color=black] (-0.2,-1.5) rectangle (0,1.5);
        
        \node[color=purple!70!white, font=\fontsize{50}{22.4}] at (2.5,0) {$\uptau$};
        
        \node[] at (2.55,1) {$\tau_b$};
        \draw[thick,->] (1.9,0.3) arc(160:20:.7);
        
        \draw[thick,->] (-.5,0) -- (4,0) node[anchor=north west] {$\!x$};
        \draw[thick,->] (0,-1.5) -- (0,1.5) node[anchor=south east] {$y$};
    \end{tikzpicture}
    \caption{A $\uptau$-shaped object generically experiences a non-zero self-torque $\tau_b$.}\label{fig:torque}
\end{figure}
\end{center}

Finally, an asymmetric object may also experience a torque from the surrounding active fluid~\cite{di_leonardo_bacterial_2010,Sokolov2010PNAS} (See Fig.~\ref{fig:torque}). In two dimensions this torque is given by
\begin{equation}\label{eq:torque}
    \boldsymbol\tau = \intop_\Omega d\vec{r}\, \rho(\vec{r}) (\vec{r}-\vec{r}_{\scriptstyle \rm CM}) \times \nabla U\;,
\end{equation}
where $\vec{r}_{\scriptstyle \rm CM}$ is the object's center of mass. Denoting the magnitude of $\boldsymbol\tau$ when $d\to\infty$ as $\tau_b$ and using the image object along with the results of Refs.~\cite{baek_generic_2018,granek2020bodies}, we find that the interaction torque $M$ due to the wall, defined through 
$\tau=\tau_b+M$, is given by
\begin{equation}
    M = \frac{\mu p_b}{4\pi D_{\rm eff}\rho_b} \frac{\cos(\phi)}{d}\tau_b + \mathcal{O}(d^{-2})\;.
\end{equation}

Note that when the object is not chiral, $\tau_b$ vanishes and there is no torque to order $\mathcal{O}(d^{-1})$. Higher order contributions are, however, expected from symmetry considerations: the density modulation along $\hat x$ due to the presence of the wall indeed breaks the chiral symmetry when $\bf p$ is not along $\hat x$.

\section{A object inside a circular cavity}\label{sec:circular cavity}

In the previous section, we showed that, far from the object and away from a boundary layer created by a flat wall, the steady-state distribution and current of active particles are equivalent to those induced by two force monopoles placed symmetrically with respect to the plane of the wall. In turn, we showed that the object interacts with its mirror image, with an interaction force given by Eq.~\eqref{eq:F12-bare}. 
We now consider a different setup of an asymmetric object placed in a circular cavity (See Fig.~\ref{fig:configuration circle}). We first determine in Section~\ref{sec:CCdens} the long-ranged density modulation and current induced by object. Then, in Section~\ref{sec:CCforce}, we compute the contribution of the force experienced by the object due to the circular confining boundary.

\begin{center}
\begin{figure}[t]
    \begin{tikzpicture}
        \clip (2-0.3,3.5) rectangle (5.5,-0.35);
        \filldraw[color=black,fill=orange!65!white] (2,0) circle (2.65);
        \filldraw[pattern=north east lines, pattern color=black] (2,0) circle (2.65);
        
        \filldraw[color=black,fill=teal!20!white] (2,0) circle (2.5);
    
        \draw[thick,->] (-1.1,0) -- (5.1,0) node[anchor=north west] {$\!x$};
        \draw[thick,->] (2,-3) -- (2,3) node[anchor=south east] {$y$};
        
        \fill[purple!70!white] (2.59611+.4,0.00139995+1) arc(158:338:0.375) -- (3.17561+.4,-0.217276+1) arc(338:158:0.25) -- cycle;
        
        \draw[-stealth,thick] (3+.3,0+.6) -- (3+.3+0.26222,0.649029+.6) node[anchor=east, xshift=-1mm] {${\bf p}$};
        \draw[-stealth,thick] (2,0) -- (3+.3,0+.6) node[anchor=south, xshift=-7mm, yshift=-3mm] {${\bf r}_0$};
        \draw[dashed,thick] (3+.3,0+.6) -- (4.2699,1.04765);
        \draw[dashed,thick] (3+.3,0+.6) -- (4.35,.6);
        \draw[line width=1.2pt,blue] (3.3 - 0.02 + 0.363184 ,.6 + 0.167623) arc(25:68:0.4) node[anchor=west, xshift=1.5mm, yshift=1.5mm] {$\psi\; $};
        
        \draw[thick] (3.3+.3 ,.6) arc(0:68:0.3) node[anchor=west, xshift=-2mm, yshift=-5mm] {$\phi\; $};
        
        \draw[thick] (2+.7 ,0) arc(0:68:0.27) node[anchor=west, xshift=2mm, yshift=-.6mm] {$\theta_0$};

    \end{tikzpicture}
    \caption{An asymmetric passive object in an active fluid placed inside a circular cavity of radius $R$. The object is located at $\vec{r}_0$ at an angle $\theta_0$ relative to the $\boldsymbol{\hat{x}}$ axis. The corresponding force monopole $\vec{p}$  is directed along $\psi=\phi-\theta_0$ relative to $\hat {\bf r}$.}\label{fig:configuration circle}
\end{figure}
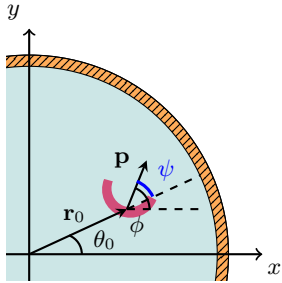
\end{center}

\subsection{Density profile and current}
\label{sec:CCdens}
Consider a passive asymmetric object placed inside an active fluid confined by a circular cavity of radius $R$. To make progress we assume that the far-field density modulation is given, to leading order, by the solution of Eq.~\eqref{eq:reduced Poisson's equation} together with the Neumann boundary condition $\hat r \cdot \nabla \rho(\bfr)|_{|\bfr|=R}=0$.
The Neumann-Green's function in this geometry can be obtained in several ways, for example, by using conformal transformations or by using the polar symmetry of the domain. It is given by~\cite{kevorkian1990partial}
\begin{align}
    G^{\rm disk}_N(\vec{r};\vec{r}_0) = -\frac{1}{2\pi}&\left[\ln\left(|\vec{r}-\vec{r}_0|/\ell_p\right)\right. \nonumber \\
    +&\left. \ln\left(|\vec{r}-\tilde{\vec{r}}_0|/\ell_p\right) + \ln\left(r_0/\ell_p\right)\right]\;,
\end{align}
with $\tilde{\vec{r}}_0 \equiv (R/r_0)^2\vec{r}_0$. Again, we write  $\rho=\rho_b + \delta\rho$, with $\rho_b$ the average density in the cavity. The leading order density modulation $\delta\rho(\vec{r})$ is then given in the far field by
\if{\begin{equation}\label{eq:density disk}
    \delta\rho \simeq - \frac{\mu}{2\pi D_{\rm eff}}\left[\left(\frac{\vec{r}-\vec{r}_0}{\left|\vec{r}-\vec{r}_0\right|^2} - \frac{\vec{r}_0}{r_0^2}\right)\cdot \vec{p} - \frac{(\vec{r}-\tilde{\vec{r}}_0)\cdot \tilde{\vec{p}}}{\left|\vec{r}-\tilde{\vec{r}}_0\right|^2}\right]\;,\nonumber
\end{equation}}\fi
\begin{equation}\label{eq:density disk}
    \delta\rho \simeq - \frac{\mu}{2\pi D_{\rm eff}}\left[\frac{(\vec{r}-\vec{r}_0)\cdot \vec{p}}{\left|\vec{r}-\vec{r}_0\right|^2} - \frac{(\vec{r}-\tilde{\vec{r}}_0)\cdot \tilde{\vec{p}}}{\left|\vec{r}-\tilde{\vec{r}}_0\right|^2}\right]+\frac{\mu \,\bfr_0\cdot \vec{p}}{2 \pi D_{\rm eff}r_0^2}\;,
\end{equation}
where $\tilde{\vec{p}}\equiv (R/r_0)^2 p \vec{u}(2\theta_0-\phi)$. The diffusive current is then obtained using $\vec{J}\simeq -D_{\rm eff}\nabla \rho$, leading to
\begin{align}
    \vec{J} \simeq \frac{\mu}{2\pi} &\left[\frac{1}{|\vec{r}-\vec{r}_0|^2}\left(\frac{2[(\vec{r}-\vec{r}_0)\cdot\vec{p}](\vec{r}-\vec{r}_0)}{|\vec{r}-\vec{r}_0|^2} - \vec{p}\right) \right. \nonumber \\
    - &\;\, \left. \frac{1}{|\vec{r}-\tilde{\vec{r}}_0|^2}\left(\frac{2[(\vec{r}-\tilde{\vec{r}}_0)\cdot\tilde{\vec{p}}](\vec{r}-\tilde{\vec{r}}_0)}{|\vec{r}-\tilde{\vec{r}}_0|^2} - \tilde{\vec{p}}\right) \right]\;.\label{eq:Jcircular}
\end{align}
Again, the current in Eq.~\eqref{eq:Jcircular} is equivalent to that generated by the force monopole and an image monopole $\tilde {\bf p}$  placed at $\tilde \bfr_0$. The same applies to the density modulation, which also experiences an additional uniform contribution that enforces mass conservation.

\begin{figure}[h!]
 	\centering
	\includegraphics[width=8.6 cm]{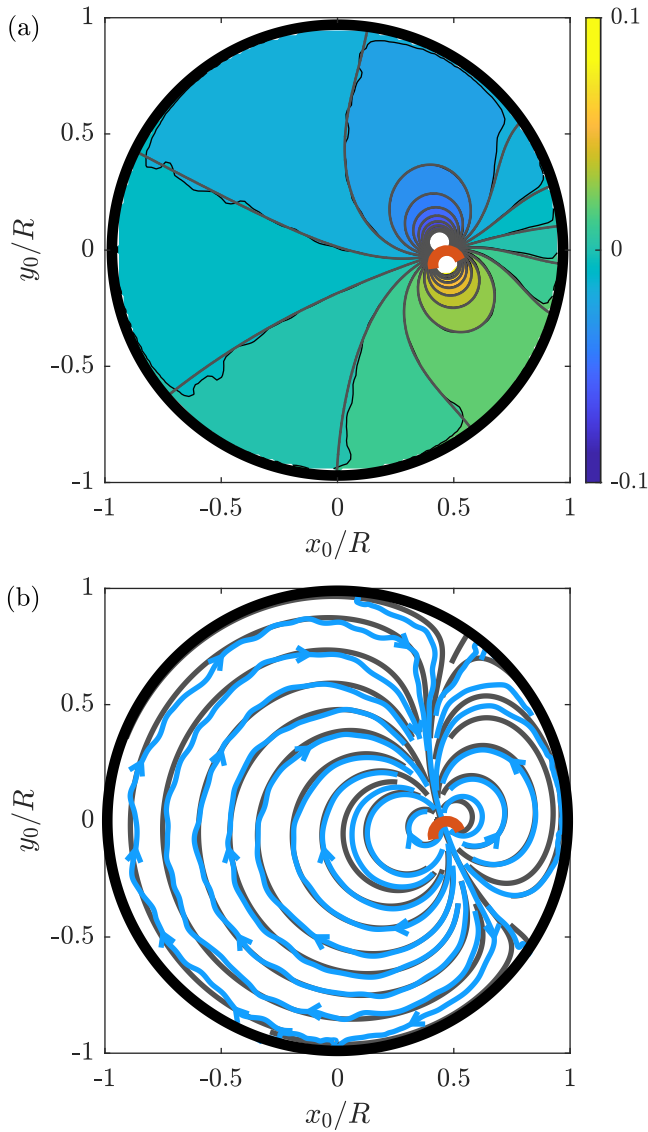}
	\caption {Density and current profiles surrounding an asymmetric object inside a circular cavity. The object, shaped as a semicircular arc of diameter $d_{\rm arc}=\ell_p$, is located at $\vec{r}_0=(0.45R,0)$ with an orientation making an angle $\phi=0.6\pi$ with the $\boldsymbol{\hat{x}}$ axis. 
	The object is displayed in orange and enlarged by a factor of six.
	(a) Steady-state density modulation relative to the bulk density, $\delta\rho/\rho_b$, compared with the analytical expression~\eqref{eq:density disk} in gray. (b) Streamlines of the steady-state current. The measurement (in light blue) is compared with the theory (in gray), for the same parameters as in (a).
    In both panels, the parameters were set as follows:  $N=10^5$ RTPs travel with speed $v=10^{-4}$ and tumble at rate $\alpha = 10^{-2}$. 
    See Appendix~\ref{app:numerics} for details.}
 	\label{fig:density and current}
\end{figure}

We verified our predictions using numerical simulations of RTPs which are shown in Figure~\ref{fig:density and current}. Both density modulations and currents  are well described by Eqs.~\eqref{eq:density disk} and~\eqref{eq:Jcircular}. 

\subsection{Interaction force}\label{sec:CCforce}
\begin{figure}
 	\centering
	\includegraphics[width=8.6 cm]{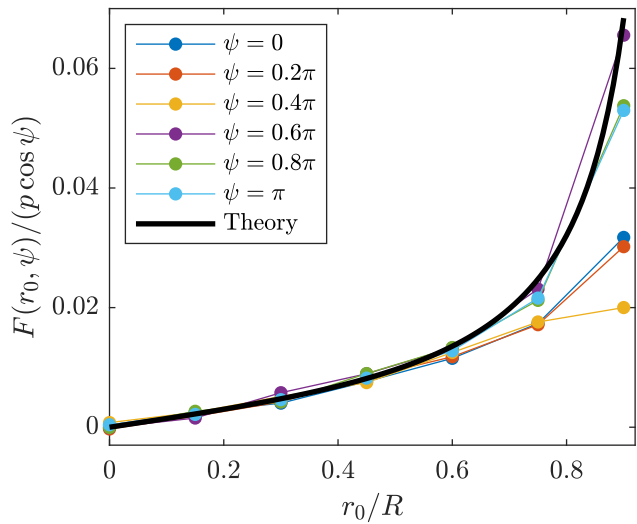}
	\caption{Collapse of the interaction force exerted on the object. The data displayed here shows magnitude of the interaction force~\eqref{eq:force circular cavity} divided by its angular dependence $\cos(\psi)=\cos(\phi-\theta_0)$ relative to the strength of the force monopole $p$. 
	The solid black line shows the theoretical prediction with no parameter fitting.
	Note that the deviation from the theory near the walls is expected, due to higher-order interactions.
	}
 	\label{fig:force collapse}
\end{figure}

Next we turn to derive the force induced on the object by the circular wall. To do this we first note that the presence of the wall leads to a density modulation
\begin{equation}
    \Delta\rho(\vec{r}_0) \approx \frac{\mu}{2\pi D_{\rm eff}}\left.\left[ \frac{\vec{r}_0}{r_0^2}\cdot \vec{p} + \frac{(\vec{r}-\tilde{\vec{r}}_0)\cdot \tilde{\vec{p}}}{\left|\vec{r}-\tilde{\vec{r}}_0\right|^2}\right]\right|_{\vec{r}=\vec{r_0}}\;,
\end{equation}
when compared to the situation in an infinite space. The force due to the presence of the wall is then given by  Eq.~\eqref{eq:F12}, which leads to:
\begin{equation}\label{eq:force circular cavity}
    \vec{F} \approx -\frac{\mu p_b^2}{2\pi D_{\rm eff}\rho_b}\frac{r_0\cos(\phi-\theta_0)}{R^2-r_0^2}\binom{\cos(\phi)}{\sin(\phi)}\;,
\end{equation}
with $p_b$ the magnitude of the force monopole measured either in the center of the cavity or equivalently for $\tilde{ r}_0 \to \infty$. As in the case of a flat wall, the force always repels the object away from the wall, as can be seen by setting $\theta_0=0$. Figure~\ref{fig:force collapse} shows a collapse of the force measured on the object for various orientations and distances from the wall, showing good agreement with the theory~\eqref{eq:force circular cavity}.

Finally, as in the case of the flat wall, we can compute the interaction torque $M$ acting on the object, which is given by
\begin{equation}
    M \approx \frac{\mu p_b}{2\pi D_{\rm eff}\rho_b}\frac{r_0\cos(\phi-\theta_0)}{R^2-r_0^2}\tau_b\;,
\end{equation}
where $\tau_b$ is the object's self-torque measured at $r_0=0$.

\section{Dynamics of an asymmetric object inside a circular cavity}\label{sec:dynamics}

In the previous section, we computed  the density modulation and current induced by a polar object held fixed in a circular cavity. The presence of confining walls leads to a renormalization of the force felt by the object which depends on its position and orientation. When the object is mobile, it is thus endowed with a non-uniform propulsion force. In this section we use a toy model to capture the corresponding dynamics and characterize its steady-state distribution. We find that the interaction with the wall leads to a transition between two distinct behaviors: the object is  localized either in the center of the cavity or near the edges, as observed in Fig.~\ref{fig:body prob dist}. 

To lighten the notations, we drop the subscript "0" when refering to the object so that its position reads ${\bf r}=r {\bf u}(\theta)$ and its orientation makes an angle $\phi$ with $\hat x$. We model the object's dynamics as an effective Langevin equation:
\begin{subequations}\label{eq:body Langevin}
\begin{align}
    \dot{\vec{r}} =& \mu_0 p \vec{u}(\phi) + \mu_0 \vec{F}(r,\theta,\phi)  + \sqrt{2D_t^e}\boldsymbol{\eta}(t) \;,\label{eq:dynpos}  \\
    \dot{\phi} =& \sqrt{2D_r^e} \xi(t)\;,
\end{align}
\end{subequations}
where $\mu_0$ is the mobility of the object, $p$ is the magnitude of $-{\bf p}_b$, $D_t^e$ and $D_r^e$ are effective translational and rotational diffusivities, and $\eta_i(t)$ and $\xi(t)$ are Gaussian white noises of zero mean and unit variance. For simplicity, we consider a symmetric object whose self-torque is zero. 

We now use the explicit expression of ${\bf F}$ given in Eq.~\eqref{eq:force circular cavity} and the angle $\psi=\phi-\theta$ between the object and $\hat {\bf r}$ (see Fig.~\ref{fig:configuration circle}) to rewrite  Eq.~\eqref{eq:body Langevin} as a dynamics for $r$, $\theta$ and $\psi$. Since $r={\bf r}\cdot \hat {r}$, It\=o calculus implies that $\dot r=\dot{\bf r}\cdot \hat r + D_t^e/r$. Similar to the case of a particle in a harmonic well~\cite{solon_active_2015}, the equations for $r$ and $\psi$ decouple from the dynamics of $\theta$, and read:
\begin{subequations}\label{eq:body Langevin 2}
\begin{align}
    \dot r =& \mu_0 p \cos\psi \left[1 -\frac{q r R \cos(\psi)}{ R^2-r^2}\right]+\frac{D_t^e}{r}+ \sqrt{2D_t^e} \eta_r(t) \;,\label{eq:dynposr}  \\
    \dot{\psi} =& -\frac{\mu_0 p \sin\psi}{r} \left[1 -\frac{q r R \cos(\psi)}{ R^2-r^2}\right]\nonumber \\ &\quad+\sqrt{2\left(\frac{D_t^e}{r^2}+D_r^e\right)} \xi_\psi(t)\;\label{eq:dynpospsi},
\end{align}
\end{subequations}
where $\eta_r$ and $\xi_\psi$ are Gaussian white noises of zero mean and unit variance and $q=\mu p/(2 \pi D_{\rm eff} \rho_b R)$ is a dimensionless parameter.

\if{
This Langevin dynamics is equivalent to the Fokker-Planck equation:
\begin{subequations}\label{eq:FP object}
\begin{align}
    \partial_t P =& - \frac{1}{r_0}\frac{\partial}{\partial r_0}\left(r_0 J_r\right) - \frac{1}{r_0}\frac{\partial J_\psi}{\partial \psi}\;,\\
    J_r =& \mu_0 p \cos(\psi) \left(1-\frac{q R^2 r_0 \cos(\psi)}{R^2 - r_0^2}\right) P - D_t^e \partial_{r_0} P\;,\\
    J_\psi =& -\mu_0 p \sin(\psi) \left(1-\frac{q R^2 r_0 \cos(\psi)}{R^2 - r_0^2}\right)P \nonumber \\
    &- \left(\frac{D_t^e}{r_0} + r_0 D_r^e\right) \partial_\psi P\;.
\end{align}
\end{subequations}
Here $J_r$ is the radial current flowing in the system, $J_\psi$ the current associated with the relative angle $\psi$, and $q\equiv \mu p/(2\pi D_{\rm eff}\rho_b R^2)$ an inverse length scale associated with the interaction with the walls. It should further be noted that the probability distribution $P(r_0,\psi)$ is normalized such that $\int dr_0 d\psi\,r_0 P(r_0,\psi)=1$.}\fi

Solving for the steady-state probability distribution $P(r,\psi)$ remains a hard task. Instead, we study the dynamics~\eqref{eq:body Langevin 2} in two limits: first when the object reorients so quickly that it is no longer persistent, $D_r^e\to\infty$, resulting in an effective equilibrium dynamics; and second, in the opposite limit, $D_r^e\to0$, when the object is highly persistent. These two regimes lead to very different behaviors that explain the transition observed in Fig.~\ref{fig:body prob dist}.

\begin{figure}
 	\centering
	\includegraphics[width=8.6 cm]{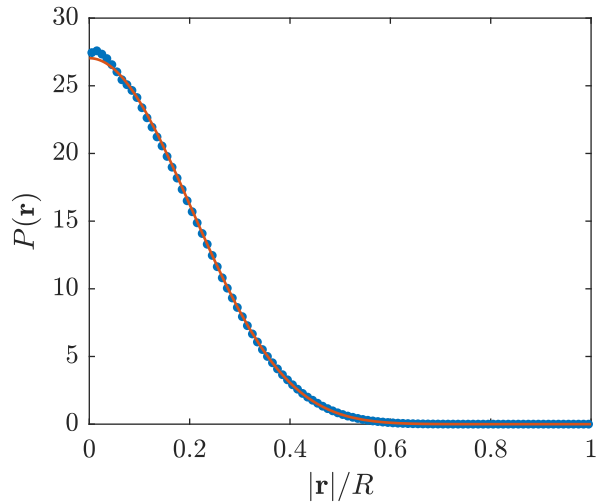}
	\caption{Steady-state probability distribution $P({\bf r})$ of the dynamics~\eqref{eq:body Langevin 2} in the large $D_r^e$ limit. Direct simulations of the Langevin dynamics~\eqref{eq:body Langevin 2} (blue dots) agree perfectly with the analytic prediction of Eq.~\eqref{eq:P diff} (orange solid line).
	}
 	\label{fig:diffusive P}
\end{figure}

\textit{Effective equilibrium limit.}
In the large $D_r^e$ limit, the dynamics of $\psi$ is dominated by the rotational diffusion, which leads to $P(r,\psi)\simeq P(r)/(2\pi)$. Taking the average of Eq.~\eqref{eq:dynposr} with respect to $\psi$ then leads to:
\begin{equation}\label{eq:body Langevin 4}
\dot r = - \frac{ \mu_0 p  q r R}{2( R^2-r^2)}+ \frac{D_t^e}{r}+\sqrt{2D_t^e} \eta_r(t) \;.
\end{equation}
The steady-state distribution of $r$ is then given by
\begin{equation}\label{eq:P diff scalar}
 P(r)\propto r \left(1-\frac{r^2}{R^2}\right)^{\frac{\mu_0 p q R}{4 D_t^e}}\;,
\end{equation}
where $P(r)$ is normalized as $\int_0^R dr\, P(r)=1$. Going back to the original ${\bf r}$ variable, one thus gets
\begin{equation}\label{eq:P diff}
    P({\bf r})\propto  \left(1-\frac{|{\bf r}|^2}{R^2}\right)^{\frac{\mu_0 p q R}{4 D_t^e}}\;,
\end{equation}
whose normalization in two dimensions reads $\int d{\bf r}\,P({\bf r})=1$. Importantly, as can be seen in Fig.~\ref{fig:diffusive P}, the distribution is peaked around ${\bf r}=0$ and perfectly matches microscopic simulations of Eq.~\eqref{eq:body Langevin 2}. The result is reminiscent of the steady-state distribution of a run-and-tumble particle in a harmonic well in one space dimension~\cite{Tailleur2008PRL,dhar2019run}.

Finally, we note that this effective equilibrium regime allows for the localization of the object in the bulk of a non-equilibrium diffusive fluid. As mentioned in the introduction, this would be impossible in equilibrium due to Earnshaw's theorem. Here, when going from Eq.~\eqref{eq:body Langevin 2} to Eq.~\eqref{eq:body Langevin 4}, the `bare' self-propulsion force of the object has cancelled out and we are only left with the contribution from its image. The reason why the latter does not lead to a vanishing contribution is the strong anti-correlation between ${\bf p}$ and its image.

\textit{Large-persistence regime}.
We now consider the opposite limit of a very small rotational diffusivity and set, for simplicity, $D_t^e=0$. In this limit, the dynamics~\eqref{eq:body Langevin 2} reduce to:
\begin{subequations}\label{eq:body Langevin 3}
\begin{align}
    \dot r =& \mu_0 p \cos\psi \left[1 -\frac{q r R \cos(\psi)}{ R^2-r^2}\right]\;,  \\
    \dot{\psi} =& -\frac{\mu_0 p \sin\psi}{r} \left[1 -\frac{q r R \cos(\psi)}{ R^2-r^2}\right]\;.\label{eq:psi dynamics persistent}
\end{align}
\end{subequations}
Following~\cite{solon_active_2015}, we expect that, in this noiseless limit, the object's position and orientation remain close to the stable fixed points of the dynamics~\eqref{eq:body Langevin 3}, found by requiring $\partial_t r=\partial_t \psi = 0$.
Direct inspection shows that all fixed points $(r^*,\psi^*)$ satisfy
\begin{equation}
    1 -\frac{q r^* R \cos(\psi^*)}{ R^2-(r^*)^2} = 0\;.
\end{equation}
There is thus a continuous line of fixed points, which can be parameterized as $r^*=r^*(\psi^*)$:
\begin{equation}\label{eq:r*}
    r^*(\psi^*) = \frac R 2 \left(\sqrt{(q \cos\psi^*)^2+4} - q \cos\psi^*\right)\;,
\end{equation}
with $\psi^*\in[-\pi/2,\pi/2]$. The minimal value of $r^*(\psi^*)$ corresponds to $\psi^*=0$ and $r^*(0)=\frac{R}2(\sqrt{q^2+4}-q)>0$. This demonstrates that, in the steady state, the object is positioned at a finite distance from the origin, unlike in the effective equilibrium limit. By changing the rotational diffusion of the object, one can thus shift its most probable localization from the center of the cavity to its periphery.
Note that Eq.~\eqref{eq:body Langevin 3} relies on the far-field approximation, which is only valid far from the walls of confining boundaries. \if{Ultimately, the only stable position of the object is at a distance close enough to the boundary where the continuum limit is no longer valid.}\fi
Ultimately, the only stable position of the object is at a distance close enough to the boundary that the modulation of the density of active particles is of order of $\rho_b$.

While the above discussion already proves the existence of the localization transition, we characterize, for completeness, the stability of the line of fixed points of the large persistence regime. As a first step, we linearize the dynamics~\eqref{eq:body Langevin 3} about $r^*(\psi^*)$. This yields a dynamical system that can be written as
\begin{equation}
    \partial_t \begin{pmatrix}
                    \delta r\\
                    \delta \psi
                \end{pmatrix} = M(\psi^*) \begin{pmatrix}
                    \delta r\\
                    \delta \psi
                \end{pmatrix} + \mathcal{O}(\delta r^2,\delta\psi^2,\delta r \delta\psi)\;,
\end{equation}  
for $\delta r\equiv r - r^*(\psi^*)$ and $\delta \psi \equiv \psi - \psi^*$, where 
\begin{subequations}
\begin{align}
    M_{11} =& -\frac{\mu_0p}{2q R}\left[4 +q \cos\psi^*\times \right. \\ 
    & \quad\times\left. \left( \sqrt{(q \cos\psi^*)^2+4}+q \cos\psi^*\right)\right]\;,\nonumber \\
    M_{12} =& \mu_0p\sin\psi^*\;,\\
    M_{21} =& \frac{mu_0p}{2 q R^2}\left[2 \tan\psi^* \sqrt{(q \cos\psi^*)^2+4}+4 q \sin\psi^*\right. \nonumber \\
    &\left.+ \frac{q^2}2 \sin(2\psi^*) \left(\sqrt{(q \cos\psi^*)^2+4}+q \cos\psi^*\right)\right]\\
    M_{22} =& -\frac{2 \mu_0 p}{R}\frac{ \sin\psi^*\tan\psi^*}{\sqrt{(q \cos\psi^*)^2+4}-q \cos\psi^*}\;.
\end{align}
\end{subequations}
Note that the matrix $M$ depends on the value of $\psi^*$. For a given value of $\psi^*$, $M$ can be diagonalized. Its eigenvectors point in two different directions: ${\bf v}_1(\psi^*)$ is tangent to the curve $r^*(\psi^*)$ and corresponds to a zero eigenvalue $\lambda_1=0$;  ${\bf v}_2(\psi^*)$ points in a different direction and is associated to a negative eigenvalue $\lambda_2$, given by
\begin{equation}
    \lambda_2(\psi^*) = -\frac{\mu_0p}{2 q R}\left(4 + q^2 + q\frac{\sqrt{(q\cos\psi^*)^2+4}}{\cos\psi^*}\right)\;.
\end{equation}
\if{It should be noted that when $\psi^*=0$ the eigenvectors point along the directions of $\delta r$ and $\delta \psi$, namely, ${\bf v}_1(\psi^*=0)=(0,1)$ and ${\bf v}_2(\psi^*=0)=(1,0)$.}\fi The direction along the line of fixed point is thus, as expected, marginally stable, whereas the transverse direction is linearly stable. To find the most probable value of $\psi^*$ and $r^*$, we thus need to go beyond the linear stability analysis and consider the non-linear dynamics of the perturbation along the line of fixed points. 

\if{The linear stability of the fixed points can be studied using the eigenvalues and the eigenvectors of $M$. Since ${\bf v}_2$ never points along $r^*(\psi^*)$ and since $\lambda_2(\psi^*)$ is always negative, the direction of ${\bf v}_2$ is stable, with a `restoring force' making the dynamics
\begin{equation}
    \partial_t {\bf v}_2 = -\lambda_2 {\bf v}_2\;.
\end{equation}
A general perturbation will thus have a component parallel to ${\bf v}_2$, that will take it back to the line $r^*(\psi^*)$.

The dynamics along the other eigenvector is different. As the eigenvector ${\bf v}_1(\psi^*)$ has a null eigenvalue, it is only marginally stable: it is not clear whether the object moves along the line $r^*(\psi^*)$, changing $\psi^*$, or stays put. To study this point, one needs to go beyond the linear stability analysis and consider the non-linear effects of the perturbation. }\fi

To do so, we expand the evolution of $\delta\psi$, given by Eq.~\eqref{eq:psi dynamics persistent}, to second order in $\delta r $ and $\delta \psi$.
Imposing a perturbation tangent to the curve $r^*(\psi^*)$ then couples $\delta r$ and $\delta \psi$. This leads to a closed dynamics for $\delta\psi$ given by
\begin{equation}
    \partial_t \delta \psi = \Gamma(\psi^*) \delta \psi^2\;,
\end{equation}
where $\Gamma(\psi^*)$ is given in Appendix~\ref{app:numerics}. Figure~\ref{fig:dtpsi} shows that $\Gamma(\psi^*)$ and $\psi^*$ have opposite signs so that $\psi^*=0$ is the sole stable fixed point.

\begin{figure}
 	\centering
	\includegraphics[width=8.6 cm]{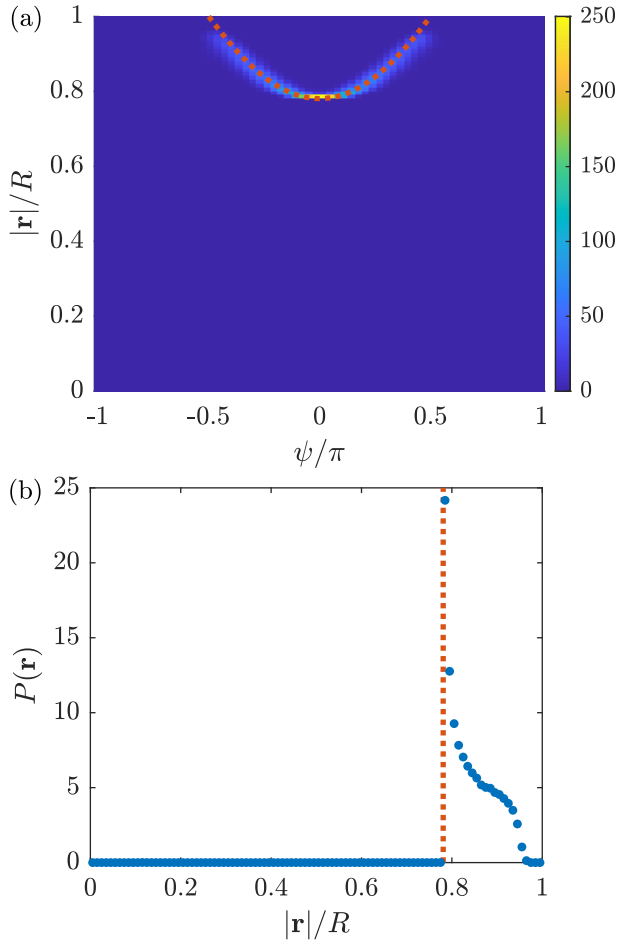}
	\caption{Steady-state behavior of the object in the small $D_r^e$ limit (with $D_t^e=0$).
	(a) The probability distribution $P({\bf r},\psi)$ measured by direct simulations of the Langevin dynamics~\eqref{eq:body Langevin 2}. The dotted orange line corresponds to the line of fixed points $r^*(\psi^*)$ given in Eq.~\eqref{eq:r*}.
	(b) The corresponding marginal probability density $P({\bf r})$. The dotted orange line marks the expected position of the object when $D_r^e=0$: $r=r^*(\psi=0)$.
	}
 	\label{fig:persistent P}
\end{figure}

In the steady state, we thus expect the object to point towards the wall, with $\psi=0$, at a distance $r=\frac{R}{2} (\sqrt{q^2+4} - q)$ from the center of the cavity. Small deviations from this solution are expected mainly along the line of fixed points $r^*(\psi^*)$. This behavior is verified by a direct numerical simulation of Eq.~\eqref{eq:body Langevin 2} presented in Fig.~\ref{fig:persistent P}.

\begin{figure}
 	\centering
	\includegraphics[width=8.6 cm]{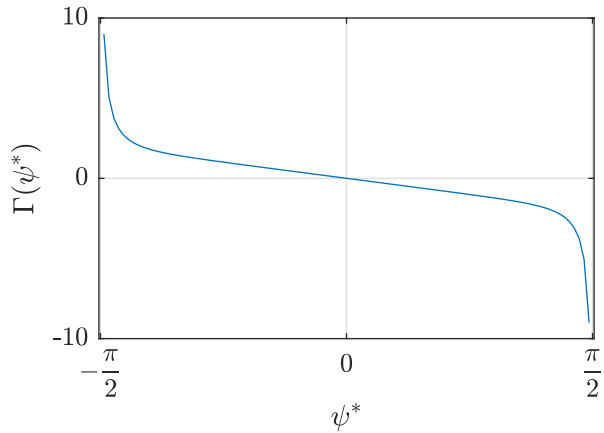}
	\caption{ $\Gamma(\psi^*)$ as a function of $\psi^*$, for $q=0.5$. The exact expression is given in Appendix~\ref{app:numerics}. For $\psi^*>0$, any perturbation $\delta\psi$ causes $\psi^*$ to decrease. For $\psi^*<0$, the opposite happens, leading to $\psi^*=0$ as the sole stable fixed point.
	}
 	\label{fig:dtpsi}
\end{figure}

\if{The dynamics near the stable fixed point $(r^*,\psi^*)=\left(\frac R 2 \left[\sqrt{q^2+4} - q\right],0\right)$ can be further described using a time-dependent Landau approach.}\fi

\begin{figure}
 	\centering
	\includegraphics[width=8.6 cm]{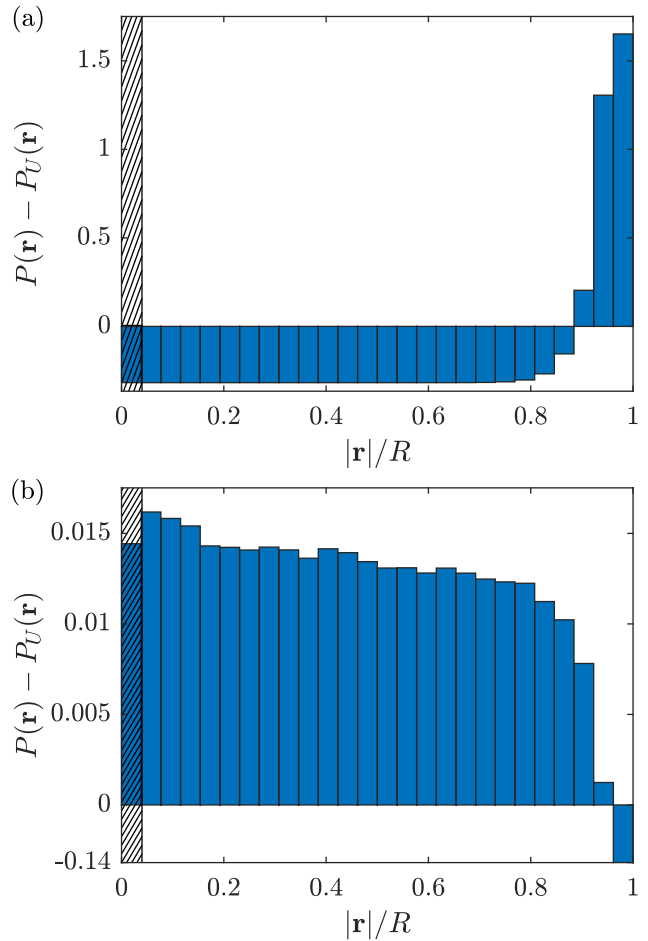}
	\caption{Comparison between the probability density of the object's position $P(\bfr)$ and a uniform distribution $P_U(\bfr)=1/(\pi R^2)$ distribution. The two panels differ by the object's rotational mobility  $\gamma=10^2$ in (a) and $\gamma=10^5$ in (b). The panels correspond to Figs.~\ref{fig:body prob dist} (a) and (b), respectively. The shaded region indicates scales smaller than the object's diameter, where the sampling is expected to fail. See Appendix~\ref{app:numerics} for numerical details.
	}
 	\label{fig:body prob dist2}
\end{figure}

All in all, the two limits of large and small rotational diffusivity show that there is a localization transition from a distribution where the object is localized close to the walls of the cavity to a distribution where the object is localized in its center. The latter occurs when the rotational diffusivity is large. 
Figure~\ref{fig:body prob dist2} shows the probability distributions of a polar object in a bath of active particles, measured numerically for the two regimes illustrated in Fig.~\ref{fig:body prob dist}, which indeed exhibits the corresponding transition.

\if{Importantly the distribution is either concave or convex near the origin depending on whether the object's persistence time $1/D_r^e$ is larger or smaller than a characteristic time $1/(2\mu_0 p q)$, which is controlled by the active bath, the object's shape and the size of the cavity. 
Figure~\ref{fig:body prob dist2} shows the probability distributions measured numerically for the two regimes that were illustrated in Fig.~\ref{fig:body prob dist}.

The underlying physics can be qualitatively understood by inspection of the Langevin dynamics~\eqref{eq:body Langevin}. In the limit of fast rotational diffusion, the `bare' self propulsion $p {\bf u}(\phi)$ averages out, while the repulsion from the wall does not. This effectively localizes the object in the center of the cavity. On the contrary, when the object is very persistent and ${\rm Pe}\ll1$, the weak correction ${\bf F}$ to the self-propulsion does not prevent the object from reaching the wall.
}\fi

\section{Conclusion}

In this manuscript we have considered the influence of boundaries on the motion of an asymmetric tracer in an active bath. Specifically, we have shown that the tracer experiences a non-conservative force mediated by the active medium,  whose magnitude depends on the object's orientation. We then demonstrated how this force can be used to control the position of the object far in the bulk of the system.

To leading order, we have shown that the interaction with the walls can be accounted for using a generalized image theorem, which states that the passive object experiences long-range forces from its image. This holds despite the non-trivial boundary condition for the active fluid near the boundary. Using this result, we showed that, inside a circular cavity, two regimes can be observed depending on the parameters: either the object is confined in the center of the cavity or it is localized close to its boundaries. All the results above are in sharp contrast to the case of a symmetric object where no stable minima can be found. 

From a broader point of view, numerous mechanisms were suggested in the past to localize objects in the center of closed domains. This was studied extensively~\cite{ierushalmi2020centering,wuhr2009does,mitchison2012growth,xie2020cytoskeleton}, in particular in the context of cell division~\cite{grill2005spindle,gundersen2013nuclear}.
Our work offers a new, simple and generic mechanism to localize a passive object in the center of a circular region without requiring any exotic interactions. While simplistic in nature, this robust mechanism might play a role in such processes. It is tempting to search for additional applications of these forces, for instance to engineer passive objects that could be controlled by modifying the boundaries of the confining system. 
Finally, it would be interesting to generalize our approach to domains of arbitrary shapes. The only non-trivial step seems to be the calculation of the Green's function, whose form can be derived using conformal mappings. 
This, however, is harder than it might seem since the Neumann boundary condition may introduce fictitious sources through the conformal mapping. A full generalization of our approach to general domains thus remains an open challenging problem.

\acknowledgements{}

YBD and YK are supported by an Israel Science Foundation grant (2038/21). YBD, YK and MK are supported by an NSF5-BSF Grant No. DMR-170828008. JT acknowledges the financial support of ANR grant THEMA.

\appendix

\section{Numerical methods}
\label{app:numerics}
\subsection{Numerical simulations of run-and-tumble dynamics with fixed obstacle}
\begin{figure}
\includegraphics[width=6 cm]{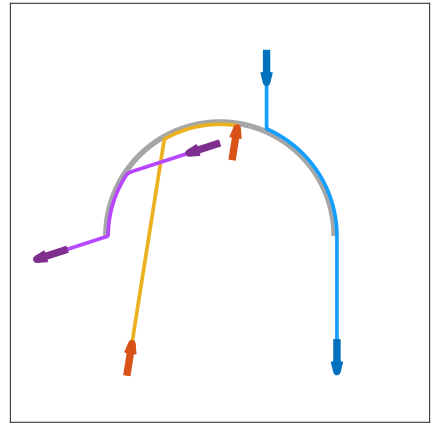}
	\caption{Trajectories of active particles modified by interactions, computed using the simulation. For clarity, the trajectories disregard tumble events, and so the particles are assumed to have constant velocities.
	A particle colliding with the outer, convex side of the semicircle (in blue) slides along the semicircle until it reaches its end. Afterwards, it continues with its original velocity.
	Particles colliding with the inner, concave side of the semicircle can have two different types of trajectories, depending on their orientation.
	One option (in purple) is that the particle slides along the inner side of the semicircle until it reaches its end. Then the particle continues its motion with its original velocity in the bulk of the cavity.
	Another option (in orange) is that the particle gets trapped inside the semicircle, stopping when its velocity is normal to semicircle. 
	Trajectories similar to the latter occur also when the particle collides with the outer walls of the cavity.
	}
 	\label{fig:numerics}
\end{figure}

Since we consider non-interacting particles, their dynamics can be ran sequentially. Each particle propagates ballistically between the tumbles whose occurrence are drawn from an exponential distribution. The dynamics of the particle is time-stepped and we use $dt=1$ and $v=10^{-4}$. The collisions with walls and obstacles are then resolved exactly: when a particle displacement leads to a collision, the dynamics is integrated until the collision, and the particle then follows the boundary tangentially until the end of the time step or of the collision. Since we use (semi-)circular shapes for boundaries, this can be done analytically. Figure~\ref{fig:numerics} illustrates the interactions of the particles with the object and the walls and shows typical trajectories obtained from the simulation. The force and the torque exerted by the particles on the object during their motion can also be derived analytically. For the force, the velocity component normal to the object is integrated over the duration of the particles' sliding motion. 

\subsection{Numerical simulations of run-and-tumble dynamics with moving object}
The dynamics of the run-and-tumble particles are now computed in parallel. We also determine the torque exerted by the particles and set the object's axis of rotation to be at the midpoint between the two arc ends. In addition, the object now moves according to 
\begin{align}
    \vec{r}(t+dt) &= \vec{r}(t) + \mu_0\overline{\vec{f} dt}\;,\nonumber\\
    \phi(t+ dt) &= \phi(t) + \gamma \overline{\tau dt}\;,
\end{align}
where ${\bf r}$ and $\phi$ are the positions and orientations of the object, and the integrated force and torques exerted by the particles during the time step $d t$ are noted $\overline{\vec{f} dt}$ and  $\overline{\tau dt}$, respectively.

\subsection{Figures~\ref{fig:body prob dist} \&~\ref{fig:body prob dist2}}
To construct Fig.~\ref{fig:body prob dist}, the object's position was measured every $10^3$ time-steps, binned in a $10^3\times10^3$ histogram, which was averaged over 500 realizations. The data were then smoothed using a Gaussian filter of width $\sigma_G = 1.5\cdot10^{-2} R$ with $R$ the radius of the cavity. Typical trajectories of the objects (in black, enlarged by a factor of 6 for legibility) are displayed in gray.
In Fig.~\ref{fig:body prob dist}(a), the simulations ran for $t_f=5\cdot10^7$. In Fig.~\ref{fig:body prob dist}(b), the simulations ran for $t_f=2.2\cdot10^8$.

Figure~\ref{fig:body prob dist2} shows the radial probability distribution corresponding to the measurement shown in Fig.~\ref{fig:body prob dist}.
The Cartesian two-dimensional distribution $P(x,y)$ was averaged over rings $r\le \sqrt{x^2+y^2} < (r+\Delta r)$ of constant thickness $\Delta r \approx 4\cdot 10^{-2} R = 2 \ell_p$. Below the scale set by the object size we encounter standard under-sampling problems due to the polar coordinates.

\subsection{Figures \ref{fig:density and current} \&~\ref{fig:force collapse}}

The data shown in the figures were averaged over 10 realizations, each running for a duration of $t_f=5\cdot10^{7}$. All measurements were carried out after an initial transient of $t=10^6$ time steps. Both figures were obtained using the same parameters.

To construct Fig.~\ref{fig:density and current}, the active particles' positions were stored and binned in a $400\times400$ histogram. In addition, their displacements during each time-step, divided by the time-step duration $d t$, were logged separately for the displacements along $\boldsymbol{\hat{x}}$ and $\boldsymbol{\hat{y}}$ on similar arrays. The data were then smoothed using a Gaussian filter of width $\sigma_G = 1.5\cdot10^{-2} R$ with $R$ the radius of the cavity.

In Fig.~\ref{fig:density and current}(a), the density field was then fitted using Eq.~\eqref{eq:density disk} to extract the bulk density $\rho_b$, the position of the equivalent force monopole $\vec{r}_0$ and its value ${\bf p}$. The figure shows the measured density relative to the fitted bulk density $\rho_b$, in units of $\rho_b$. Gray and black solid lines show comparisons between predicted and measured iso-density lines. For clarity, the object is shown in orange, enlarged by a factor of 6.

Figure~\ref{fig:density and current}(b) shows streamlines of the measured current density in blue obtained using the Matlab tool streamslice. Using the measured strength and position of the force monopole we compare them with the predictions from Eq.~\eqref{eq:density disk} in gray.

In Fig.~\ref{fig:force collapse}, to measure the force ${\bf F}$ induced by the wall on the object, we first measured ${p_b}$ by time-averaging the force exerted on an object placed at ${\bf r}_0=0$, where ${\bf F}=0$, by symmetry.

We then measure the total force $-{\bf p}$ exerted on objects placed at various distances $r$ from the center of the cavity, with different orientations $\psi$. We then added ${\bf p}_b$ to this total force to extract the contribution ${\bf F}$ from the wall, using Eq.~\eqref{eq:force circular cavity}.

Data measured for different values of $\psi$ were collapsed by dividing ${\bf F}$ by $\cos \psi$. Furthermore, the data are normalized by $p_b$. Our measurement of $p_b$ agrees quantitatively with the fit done in Fig.~\ref{fig:density and current}. Finally, the solid black curve shows the dependence according to Eq.~\eqref{eq:force circular cavity}, obtained using the measurements of $\mu p$ and $\rho_b$, measured in Fig.~\ref{fig:density and current}(a). 

\subsection{Figures~\ref{fig:diffusive P} \&~\ref{fig:persistent P} }
To produce the figures, the Langevin dynamics~\eqref{eq:body Langevin 2} were simulated using Euler time stepping with $dt=10^{-1}$. 

For Fig.~\ref{fig:diffusive P}, the steady-state distribution was obtained by binning $r$ in 100 bins and dividing by the Jacobian $r$ to get $P({\bf r})$. The latter was then compared in the Figure with the prediction of Eq.~\eqref{eq:P diff}. Parameters: $\mu_0 p=10^{-4}$, $D_t^e=10^{-6}$, $D_r^e=10$, $q=0.5$. Data were averaged over a total time of $t_f=10^9$.

For Fig.~\ref{fig:persistent P}, the steady-state distribution $P({\bf r},\psi)$ was done by binning the values of both $r$ and $\psi$ in a $100\times 64$ array. $P$ was again divided by the Jacobian $r$. Parameters: $\mu_0 p=10^{-4}$, $D_r^e=10^{-5}$, $q=0.5$. The simulation was carried for a duration of $t_f=10^8$. 

\subsection{Figure~\ref{fig:dtpsi}}
The dynamics of $\delta \psi$ along the line of fixed points is given, to second order in $\delta \psi$ by:
\begin{align}\label{eq:expression dtpsi}
    \partial_t \delta\psi \equiv& \Gamma(\psi^*)\delta\psi^2 \nonumber \\
    =&\frac{\delta \psi^2}{4 R} \tan \psi^*\left[2 q(64 + 38q^2 +5q^4) \right. \nonumber \\
    &\quad + q (256 + 112 q^2 +15q^4 ) \cos\psi^*\nonumber \\
    &\quad +6q^3 (6 + q^2) \cos(4\psi^*) + q^5 \cos(6\psi^*)\nonumber \\
    &\quad -4\sqrt{(q \cos\psi^*)^2+4}\cos\psi^*
    \left\{32 +3q^2(4+q)\right.\nonumber \\
    &\left.\left.\qquad + 4q^2 (7+q^2)\cos(2\psi^*) + q^4 \cos(4\psi^*)\right\}\right] \nonumber \\
    &\times\left[(q \cos\psi^*)^2+4\right]^{-1} \nonumber \\
    &\times\left[\sqrt{(q \cos\psi^*)^2+4} -q \cos\psi^* \right]^{-4}\;.
\end{align}

\section{Far-field density profile of pairwise-interacting active particles}\label{app:PFAPs}

In this Appendix, we generalize the proof of Section~\ref{sec:flat wall derivation} to pairwise-interacting particles, showing they exhibit  density and current profiles with the same functional form as non-interacting particles in the far field of both the object and the wall.

We start by considering the Langevin dynamics of the $i^{\rm th}$ active particle, located at $\bfr_i$ with an orientation $\theta_i$
\begin{subequations}\label{eq:interacting dynamics}
\begin{align}
    \frac{d{\bf r}_i}{dt} =& v\vec{u}(\theta_i) - \mu \nabla \Big[U({\bf r}_i) + \sum_{j\ne i} \mathfrak{u}(|{\bf r}_i-{\bf r}_j|)\Big] \nonumber \\
    & +\sqrt{2D_t} \boldsymbol{\eta}_i\left(t\right)\ , \\
    \frac{d\theta_i}{dt} =& \sqrt{2D_r}\xi_i(t)\ .
\end{align}
\end{subequations}
Here, on top of the dynamics in the dilute limit leading to Eq.~\eqref{eq:FP}, we add an isotropic inter-particle interaction potential $\mathfrak{u}(|\bfr_i-\bfr_j|)$. The particle's motion is subject to translational and rotational noises, $\xi_i(t)$ and $\boldsymbol{\eta}_i(t)$ respectively, which are assumed to be Gaussian, centered and of unit variances.

Using It\=o calculus, 
one can derive the evolution equation of the empirical distribution $\Pi(\bfr,\theta,t) \equiv \sum_i \delta^{(2)}(\bfr-\bfr_i)\delta(\theta-\theta_i)$ of the active particles starting from dynamics~\eqref{eq:interacting dynamics}~\cite{farrell2012pattern}. Then, the evolution of the mean density $\rho(\bfr,t)\equiv\langle \hat{\rho}(\bfr,t) \rangle = \langle\sum_i\delta^{(2)}(\bfr-\bfr_i)\rangle$ can be derived, leading to a continuity equation in the steady state, similarly to Eqs.~\eqref{eq:rho}-\eqref{eq:sigmas} in the dilute limit~\cite{solon_generalized_2018,solon_pressure_2015-3,granek2020bodies}:
\begin{subequations}
\begin{align}
    \partial_t \rho =& -\nabla\cdot\vec{J}=0\;, \\
    \vec{J} =& -\mu\rho\nabla U  + \mu \nabla\cdot\sigma \;, \\
    \sigma_{ij} =& -\frac{D_{\rm eff}}{\mu}\rho\delta_{ij} + \Sigma_{ij} + \sigma^{\rm P}_{ij} + \sigma^{\rm IK}_{ij}\;.\label{eq:sigma def}
\end{align}
\end{subequations}
Here $\boldsymbol{\Sigma}$ reads:
\begin{equation}
    \Sigma_{ij} = -\frac{v\tau }{\mu}\left[v Q_{ij}-\left(\mu \partial_j U + D_t\partial_j\right) m_i\right]\;,
\end{equation}
where $m_k(\bfr,t) \equiv \langle \hat{m}_k (\bfr,t) \rangle = \langle \sum_i u_k(\theta)\delta^{(2)}(\bfr-\bfr_i)\rangle$ and 
$Q_{ij}(\bfr,t) \equiv \langle \sum_k \left[u_i(\theta)u_j(\theta)-\delta_{i,j}/2\right]\delta^{(2)}(\bfr-\bfr_k)\rangle$,
$\boldsymbol{\sigma}^{\rm P}$ is the tensor
\if{
\begin{align}\label{eq:sigma P}
	\boldsymbol{\sigma}^{\rm P}\left({\bf r}\right) =& \ell_p \intop d^2 r'\,\nabla \mathfrak{u}\left(\left|{\bf r}-{\bf r}'\right|\right) \langle \hat{{\bf m}} \left({\bf r}\right) \hat{\rho}\left({\bf r}'\right)\rangle \nonumber \\
	& + \frac{D_t\ell_p}{\mu}\nabla{\bf m}\left({\bf r}\right) - \frac{v\ell_p}{\mu}{\bf Q}\left({\bf r}\right)\ ,
\end{align} 
}\fi
\begin{align}\label{eq:sigma P}
	\boldsymbol{\sigma}^{\rm P}\left({\bf r}\right) =& \ell_p \intop d^2 r'\,\nabla \mathfrak{u}\left(\left|{\bf r}-{\bf r}'\right|\right) \langle \hat{{\bf m}} \left({\bf r}\right) \hat{\rho}\left({\bf r}'\right)\rangle\ ,
\end{align} 
and $\boldsymbol{\sigma}^{\rm IK}$ is the Irving-Kirkwood stress~\cite{irving1950statistical,kruger2018stresses}
\begin{align}
	\boldsymbol{\sigma}^{\rm IK}\left({\bf r}\right) =& \frac{1}{2}\intop d^2 r'\, \frac{{\bf r}'{\bf r}'}{\left| {\bf r}' \right|} \mathfrak{u}'\left({\bf r'}\right) \\
	&\times\intop_0^{1} d\lambda\,\langle \hat{\rho}\left({\bf r}+\left(1+\lambda\right){\bf r}'\right)\hat{\rho}\left({\bf r}+\left(1-\lambda\right){\bf r}'\right)\rangle \;.\nonumber\label{eq:sigma IK}
\end{align}

As in the dilute limit derived in Section~\ref{sec:flat wall derivation}, it is thus helpful to introduce a residual current
\begin{equation}
    \J^\sigma \equiv \vec{J} - \mu\nabla\cdot\boldsymbol{\sigma}\;.
\end{equation}
Using the steady-state equation $\nabla\cdot \vec{J}=0$ then leads to
\begin{equation}\label{eq:Poisson interacting}
    \mu \partial_i\partial_j \sigma_{ij} = -\partial_i \mathcal{J}^\sigma_i\;.
\end{equation}
Furthermore, the no-flux boundary condition across the wall leads to:
\begin{equation}
    J_x(0,y) = \left.\left(\mu \partial_j \sigma_{xj} + \mathcal{J}^\sigma_x\right)\right|_{x=0}=0\;.
\end{equation}
Summation over repeated indices is implied henceforth. From there, we generalize for the stress tensor $\boldsymbol{\sigma}$ what we did for the density field in the main text.

We first employ a Helmholtz-Hodge decomposition for the stress-tensor's divergence
\begin{equation}
    \nabla\cdot\boldsymbol{\sigma} \equiv -\nabla\Phi + \nabla\times \boldsymbol{\Psi}\;,
\end{equation}
where $\boldsymbol{\Psi} \equiv {\bf \hat{z}}\Psi$ is a vector potential and $\Phi$ -- a scalar potential. With this, the tensor equation~\eqref{eq:Poisson interacting} reduces to a Poisson equation, describing the scalar field $\Phi$
\begin{align}\label{eq:Poisson Phi}
    \mu \nabla^2 \Phi =& \nabla\cdot \J^\sigma\;, \nonumber \\
    \left.\mu \partial_x \Phi \right|_{x=0} =& \left.\left(\mathcal{J}^\sigma_x + \mu \partial_y \Psi\right)\right|_{x=0}\;,
\end{align}
similar to Eqs.~\eqref{eq:Poisson}-\eqref{eq:BC} in the dilute limit.

In the absence of the object $U=0$, the solution $\Phi\fw(\bfr)$ to the boundary-value problem~\eqref{eq:Poisson Phi} is uniform in the bulk of the system, with a finite-sized boundary layer close to the wall. We denote the quantities associated with this solution by $\Phi\fw$, $\Psi\fw$, $\J^\sigma\fw$, and $\rho\fw$.

We now use the solution $\Phi\fw(\bfr)$ to obtain the far-field behavior of the full solution $\Phi(\bfr)$ in the presence of the object. To this end, we decompose $\Phi$ as $\Phi(\bfr) \equiv \Phi\fw(\bfr) + \delta\Phi(\bfr)$ to find
\begin{subequations}\label{eq:delta Phi}
\begin{align}
    \mu\nabla^2\delta\Phi =& \nabla\cdot\delta\J^\sigma\;, \\
    \delta\mathcal{J}^\sigma_i =& -\mu\rho\partial_i U\;,\label{eq:deltaJ} \\
    \left.\mu \partial_x \delta\Phi \right|_{x=0} =& \left.\left(\delta\mathcal{J}^\sigma_x + \mu \partial_y \delta\Psi\right)\right|_{x=0}=\left.  \mu \partial_y \delta\Psi\right|_{x=0}\;.
\end{align}
\end{subequations}
Here $\delta\J^\sigma\equiv\J^\sigma-\J^\sigma\fw$ and $\delta\boldsymbol\Psi\equiv\boldsymbol\Psi-\boldsymbol\Psi\fw$, thanks to the linearity of Poisson's equation. Equations~\eqref{eq:delta Phi} describe the modulation $\delta\Phi$ created by an asymmetric object on top of the solution $\Phi\fw$ induced by the flat wall.

We can now readily use the Neumann-Green's function of the half-plane, given in Eq.~\eqref{eq:Neumann-Green's function flat wall}, to solve the boundary value problem~\eqref{eq:delta Phi}. The solution reads
\begin{align}
    \mu \delta\Phi(\vec{r}) =& -\intop_{0}^{\infty} \!\!dx' \!\!\!\intop_{-\infty}^{\infty} \!\!dy'\, G_N(x,y;x',y') \nabla'\cdot\delta\J^\sigma{}'  \nonumber \\
    & - \intop_{-\infty}^{\infty}\!\! dy'\,G_N(x,y;0,y') \left.\mu\partial_x'\delta\Phi'\right|_{x'=0} 
\end{align}
Then, using Eq.~\eqref{eq:deltaJ} and performing an integration by parts, we get
\begin{subequations}
\begin{align}
    \mu \delta\Phi(\vec{r})=& -\intop_{0}^{\infty} \!\!dx' \!\!\!\intop_{-\infty}^{\infty} \!\!dy'\, \mu \rho' \nabla' U \cdot \nabla' G_N(x,y;x',y')\label{eq:Phi int 1} \\
    & - \intop_{-\infty}^{\infty}\!\! dy'\,G_N(x,y;0,y') \label{eq:Phi int 3}\left. \mu\partial_y'\delta\Psi' \right|_{x'=0}\;.
\end{align}
\end{subequations}

We now turn to evaluate this solution in the far field of both the object and the wall, where $(\vec{r}-\vec{r}_0),d\gg a,\ell_p$. As we show below, to leading order in the far field, $\delta\Phi$ is dominated by the integral~\eqref{eq:Phi int 1} so that:
\begin{align}\label{eq:Phi solution}
    \delta\Phi(\vec{r}) \underset{|\vec{r}-\vec{r}_0|,d\gg a,\ell_p}{\simeq}&\frac{1}{2\pi}\cdot\left[\frac{\vec{p}\cdot(\vec{r}-\vec{r}_0)}{\left|\vec{r}-\vec{r}_0\right|^2} + \frac{\vec{p}^\perp\cdot(\vec{r}-\vec{r}^\perp_0)}{\left|\vec{r}-\vec{r}^\perp_0\right|^2}\right] \nonumber \\
    & + \mathcal{O}\left(\left|\vec{r}-\vec{r}_0\right|^{-2},d^{-2}\right)\;,
\end{align}
where the force monopole $\vec{p}$ has the same expression as in the dilute limit~\eqref{eq:force monopole p flat wall}. 
As detailed in Ref.~\cite{granek2020bodies}, it is possible to show that $\delta\boldsymbol{\Psi}$ is of higher order compared to $\delta\Phi$. Namely, of order $\mathcal{O}(|\bfr-\bfr_0|^{-2},d^{-2})$. Then, to leading order, 
$\sigma_{ij} \simeq \sigma_{{\rm FW},ij} + \delta_{ij}\delta\Phi$. Furthermore, in the far field of the object, the stress tensor satisfies a local equation of state:
\begin{equation}
    \boldsymbol{\sigma} (\vec{r}) =  \boldsymbol{\sigma} (\rho(\vec{r})) + \mathcal{O}(\nabla\rho)\;.
\end{equation}
Denoting $\mathcal{P}=-\frac 1 2 \textrm{Tr} \sigma$ leads to $\mathcal{P}'(\rho_b) \delta \rho = \delta \Phi$, with $\mathcal{P}'=\partial \mathcal{P}/\partial\rho$ so that:
\begin{align}\label{eq:density profile int}
    \delta\rho(\vec{r}) \underset{|\vec{r}-\vec{r}_0|,d\gg a,\ell_p}{\simeq}& \frac{\vec{p}}{2\pi \mathcal{P}'(\rho_b)}\cdot\left[\frac{\vec{r}-\vec{r}_0}{\left|\vec{r}-\vec{r}_0\right|^2} + \frac{\vec{r}^\perp-\vec{r}_0}{\left|\vec{r}^\perp-\vec{r}_0\right|^2}\right] \nonumber \\
    & + \mathcal{O}\left(\left|\vec{r}-\vec{r}_0\right|^{-2},d^{-2}\right)\;.
\end{align}

\if{
\section{Validity of the multipole expansion}\label{app:multipole}

Here we show that Eqs.~\eqref{eq:rho3}-\eqref{eq:rho4} are indeed negligible compared to~\eqref{eq:rho2} in the far field of both the object and the wall. The main difficulty of the proof stems from the non-local structure of $\delta\boldsymbol{\Sigma}$ which is non-vanishing across the entire space accessible to the active fluid. To proceed, we divide the half-plane in two regions: one, denoted $\Omega_o$, which surrounds the object, and its complement, denoted $\Omega_o^c$ (see Fig.~\ref{fig:regions}). These are chosen such that on $\Omega_o^c$ $U$ vanishes  and $\delta\boldsymbol{ m},\delta\boldsymbol{Q}$ are well-approximated by their multipole-expansion form.

\begin{center}
\begin{figure}[t]
    \begin{tikzpicture}
        \fill[mycolor1] (0,-1.5) rectangle (4,1.5) node[anchor=north east,xshift=-.5mm, yshift=-1mm, color=black] {$\Omega_o^c$};;
        \fill[mycolor2] (3,0) circle (.7) node[anchor=south west, xshift=1mm, yshift=1mm, color=black] {$\Omega_o$};
        \fill[orange!65!white] (-0.2,-1.5) rectangle (0,1.5);
        \fill[pattern=north west lines, pattern color=black] (-0.2,-1.5) rectangle (0,1.5);
        
        \draw[thick,->] (-.5,0) -- (4,0) node[anchor=north west] {$\!x$};
        \draw[thick,->] (0,-1.5) -- (0,1.5) node[anchor=south east] {$y$};
        
        \fill[purple!70!white] (2.614845-0.0124869*1.5,0.0477591-0.0309061*1.5) arc(158:-22:0.375) -- (3.19434-0.0124869*1.5,-0.18637-0.0309061*1.5) arc(-22:158:0.25) -- cycle;
        
        \draw[-stealth,thick] (0,0) -- (3,0) node[anchor=east, yshift=-3mm, xshift=2mm] {${\bf r}_0$};
    \end{tikzpicture}
    \caption{Regions $\Omega_o$, enclosing the object, and  $\Omega_o^c.$}\label{fig:regions}
\end{figure}
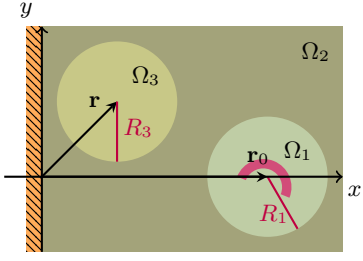
\end{center}

We start by discussing the region enclosing the object, $\Omega_o$, contributing to the integral of Eq.~\eqref{eq:rho3} but not to that of Eq.~\eqref{eq:rho4}. As $\Omega_o$ is closed and finite, one can treat $\delta\boldsymbol{\Sigma}$ in this region as a localized source density. It is thus possible to perform a multipole expansion to compute its contribution to the density modulation and current at a position ${\bf r}$ in the far field. The integral over $\Omega_0$ involving $\delta\boldsymbol{\Sigma}$ in~\eqref{eq:rho3} has more derivatives---and is hence of a higher order contribution---than the contribution of Eq.~\eqref{eq:rho2}. 

Next, we turn to the complement region $\Omega_o^c$. We remind the reader that by Eq.~\eqref{eq:Sigma} $\delta\boldsymbol{\Sigma}$ is given in $\Omega_o^c$ by
\begin{equation}
    \delta\Sigma_{ij}\bigg|_{\bfr\in\Omega_o^c} = -\frac{v\tau}{\mu}\left[v \delta Q_{ij} - D_t \partial_j \delta m_i\right]\;.
\end{equation}
By construction, in this region $\delta\boldsymbol{m}$ and $\delta\boldsymbol{Q}$ take their far-field forms. Then, Eqs.~\eqref{eq:FP} and~\eqref{eq:m} show that $\delta\boldsymbol{m}\sim \nabla\rho$ and $\delta\boldsymbol{Q}\sim \nabla\nabla\rho$. Using the density profile~\eqref{eq:density force monopole flat wall} self-consistently, we thus find
\begin{equation}\label{eq:delta Sigma scaling}
    \delta\boldsymbol{\Sigma}\bigg|_{\bfr\in\Omega_o^c} \sim \frac{1}{|\bfr-\bfr_0|^3}\;,
\end{equation}
to leading order in the far field. Note that the contributions for the image follow the same scaling.

Eq.~\eqref{eq:delta Sigma scaling} can now be inserted into Eq.~\eqref{eq:rho3}. Using the Green's function explicitly and performing an integration by parts to match the structure of the integral~\eqref{eq:rho2}, we find a contribution to $\delta\rho$ of the form
\begin{equation}\label{eq:Sigma contribution}
    \intop_{\Omega_o^c} d^2\bfr'\, \left[\partial_i'G_N(\bfr;\bfr')\right] \left[\partial_j'\delta\Sigma_{ij}'\right] \sim \intop_{\Omega_o^c} \frac{d^2\bfr'}{|\bfr-\bfr'|} \frac{1}{|\bfr'-\bfr_0|^4}\;.
\end{equation}
Since $\bfr_0$ is outside of $\Omega_o^c$, the only divergence of this expression might arise due to sources positioned at $\bfr'=\bfr$. To show that there is in fact no divergence, it is useful to define $\vec{u}\equiv\bfr-\bfr'$. With this, the expression~\eqref{eq:Sigma contribution} becomes
\begin{equation}\label{eq:integral u}
    \intop_{\Omega_o^c} \frac{u du d\gamma}{u} \frac{1}{|\bfr-\bfr_0-\vec{u}|^4}\;,
\end{equation}
with $u=|\vec{u}|$ and $\tan\gamma=u_y/u_x$, showing that the $u\to0$ limit is regular. 

The integral~\eqref{eq:integral u} can be evaluated as follows: the leading contribution comes from regions where $\vec{u}\approx\bfr-\bfr_0$. This occurs on the boundaries of $\Omega_o$ and leads to a contribution that is independent of $\bfr$. The contribution then renormalizes the bulk density $\rho_b$ and does not affect the multipole expansion.

Apart from the contribution due to this short-distance cutoff, one can see directly from the expression~\eqref{eq:integral u} that it involves a contribution scaling as $|\bfr-\bfr_0|^{-3}$. This stems from the integral over the bulk of the system. 
However, as the domain is restricted by the wall, additional dependencies on $\bfr$ arise from this integral. By considering different limits, one finds two additional contributions depending on the distance $d$ of the object from the wall.
To see this, let $\varphi$ be the angle between $(\bfr-\bfr_0)$ and ${\bf u}$. The integral~\eqref{eq:integral u} then reads
\begin{align}
    &\intop_{\Omega_o^c} \frac{du d\gamma}{\left[u^2+|\bfr-\bfr_0|^2 - 2 u |\bfr-\bfr_0| \cos(\varphi-\gamma)\right]^2} \\
    &\qquad\qquad\qquad\qquad\qquad\qquad \approx\intop_{\Omega_o^c} \frac{du}{\left[\max{\left\{u,|\bfr-\bfr_0|\right\}}\right]^4}\;.\nonumber
\end{align}
With this expression, it is easy to address the different limits. Specifically, considering the scales $|\bfr-\bfr_0|$ and $d$, one finds that the integral over region where $u>|\bfr-\bfr_0|$ adds a contribution of order $\mathcal{O}(d^{-3})$ whereas the $u<|\bfr-\bfr_0|$ region adds a contribution of order $\mathcal{O}(d/|\bfr-\bfr_0|^4)$. Except for the renormalization of the bulk density, all these contributions are indeed negligible in the far field compared with the force-monopole contribution leading to Eq.~\eqref{eq:density force monopole flat wall}.

Finally, a similar analysis can be carried out for the surface term~\eqref{eq:rho4}. The multipole expansion form of $\delta\boldsymbol{\Sigma}$ suggests that the integral involving $\partial_x'\delta\Sigma_{xx}'\big|_{x'=0}$ vanishes (by symmetry, due to the image), and so the only contribution arises from the $\partial_y'\delta\Sigma_{xy}'\big|_{x'=0}$ integral of~\eqref{eq:rho4}. Performing an integration by parts leads to an expression similar to Eq.~\eqref{eq:integral u}, which in turn reveals contributions scaling as $|y-y_0|^{-3}$ and $d^{-3}$. These contributions are negligible in the far field, meaning that one can disregard the surface term~\eqref{eq:rho4} as well.

All in all, both contributions coming from $\Omega_o$ and $\Omega_o^c$ are shown to be negligible in the far field compared to the force-monopole contribution of Eq.~\eqref{eq:rho2}, concluding the argument.
}\fi

\section{Validity of the multipole expansion}\label{app:multipole}

Here we show that Eqs.~\eqref{eq:rho3}-\eqref{eq:rho4} are indeed negligible compared to~\eqref{eq:rho2} in the far field of both the object and the wall. For legibility, we remind here the full expression of the integral:
\begin{subequations}
\begin{align}
    \delta \rho(x,y) =&   -\frac{\mu}{D_{\rm eff}}\intop_0^\infty dx'\intop_{-\infty}^\infty dy'\,\rho'\nabla'U \cdot \nabla' G_N(x,y;x',y')\label{eq:app:rho2}\\
    & -\frac{\mu}{D_{\rm eff}}\intop_0^\infty dx'\intop_{-\infty}^\infty dy'\, G_N(x,y;x',y') \partial_i'\partial_j'\delta\Sigma_{ij}'\label{eq:app:rho3} \\
    & - \frac{\mu}{D_{\rm eff}}\intop_{-\infty}^\infty dy'\,G_N(x,y;0,y')\partial_j'\delta\Sigma_{xj}'\bigg|_{x'=0}\;. \label{eq:app:rho4}
\end{align}
\end{subequations}
The main difficulty of the argument stems from the non-local structure of $\delta\boldsymbol{\Sigma}$ which is non-vanishing across the entire space accessible to the active fluid. To proceed, we divide the half-plane in three regions: $\Omega_1$ is a disk of radius $R_1$ centered on the object, $\Omega_3$ is a disk of radius $R_3$ centered on the position ${\bf r}$ where the density field is evaluated, and $\Omega_2$ is the complement to $\Omega_1 \cup \Omega_3$ (see Fig.~\ref{fig:regions}). For reasons that will become clear later on, we set $R_3=\varepsilon |\bfr-\bfr_0|$, with $\varepsilon$ a constant less than $1/2$. We also set $R_1$ to be large enough that all fields outside $\Omega_1$ can be evaluated using their far-field multipolar expressions.

\begin{center}
\begin{figure}[t]
    \begin{tikzpicture}
         \fill[mycolor1] (0,-1) rectangle (4,2) node[anchor=north east,xshift=-.5mm, yshift=-1mm, color=black] {$\Omega_2$};
        \fill[mycolor2] (3,0) circle (.8) node[anchor=south west, xshift=1mm, yshift=1mm, color=black] {$\Omega_1$};
        \fill[orange!65!white] (-0.2,-1) rectangle (0,2);
        \fill[pattern=north west lines, pattern color=black] (-0.2,-1) rectangle (0,2);
        \fill[mycolor3] (1,1) circle (.8) node[anchor=south west, xshift=.8mm, yshift=1mm, color=black] {$\Omega_3$};
        
        \fill[purple!70!white] (2.614845-0.0124869*1.5,0.0477591-0.0309061*1.5) arc(158:-22:0.375) -- (3.19434-0.0124869*1.5,-0.18637-0.0309061*1.5) arc(-22:158:0.25) -- cycle;
        
        \draw[-stealth,thick] (0,0) -- (3,0) node[anchor=east, yshift=2.5mm, xshift=1.5mm] {${\bf r}_0$};
        \draw[-stealth,thick] (0,0) -- (1,1) node[anchor=east, yshift=0mm, xshift=-1mm] {${\bf r}$};
	    \draw[thick,purple] (1,.2) -- (1,1) node[anchor=east, yshift=-3.5mm, xshift=6mm] {$R_3$};
        \draw[thick,purple] (3,0) -- (3.4,-0.693) node[anchor=east, yshift=2mm, xshift=0mm] {$R_1$};
        
        \draw[thick,->] (-.5,0) -- (4,0) node[anchor=north west] {$\!x$};
        \draw[thick,->] (0,-1) -- (0,2) node[anchor=south east] {$y$};
    \end{tikzpicture}
    \caption{Regions $\Omega_o$, enclosing the object, and  $\Omega_o^c.$}\label{fig:regions}
\end{figure}
\end{center}

Integrating by parts Eq.~\eqref{eq:app:rho3} first cancels the contribution of Eq.~\eqref{eq:app:rho4}. A second integration by parts leads to
\begin{equation}
    -\frac{\mu}{D_{\rm eff}}\intop_0^\infty dx'\intop_{-\infty}^\infty dy'\, \partial_i'\partial_j' G_N(x,y;x',y') \delta\Sigma_{ij}'\label{eq:app:rho3bis}\;.
\end{equation}
The lack of any other boundary term is due to the boundary conditions on the Green's function and on $\delta \boldsymbol{\Sigma}$.

We start by discussing the region $\Omega_1$ enclosing the object and its contribution to Eq.~\eqref{eq:app:rho3bis}. As $\Omega_1$ is closed and finite, one can treat $\delta\boldsymbol{\Sigma}$ in this region as a localized source density. It is thus possible to perform a multipole expansion to compute its contribution to the density modulation and current at a position ${\bf r}$ in the far field. The integral over $\Omega_1$ has one more derivative---and is hence of a higher order---than the contribution of Eq.~\eqref{eq:app:rho2}. 

Before we turn to the regions $\Omega_2$ and $\Omega_3$, 
 we note that Eq.~\eqref{eq:Sigma} implies that $\delta\boldsymbol{\Sigma}$ is given by
\begin{equation}
    \delta\Sigma_{ij}\bigg|_{\bfr\in\Omega_1^c} = -\frac{v\tau}{\mu}\left[v \delta Q_{ij} - D_t \partial_j \delta m_i\right]\;
\end{equation}
in the region $\Omega_1^c=\Omega_2\cup \Omega_3$.  
By construction, in this region, $\delta\boldsymbol{m}$ and $\delta\boldsymbol{Q}$ take their far-field forms. Then, Eqs.~\eqref{eq:FP} and~\eqref{eq:m} show that $\delta\boldsymbol{m}\sim \nabla\rho$ and $\delta\boldsymbol{Q}\sim \nabla\nabla\rho$. Using the density profile~\eqref{eq:density force monopole flat wall} self-consistently, we thus find
\begin{equation}\label{eq:delta Sigma scaling}
    \delta\boldsymbol{\Sigma}\bigg|_{\bfr\in\Omega_1^c} \sim \frac{1}{|\bfr-\bfr_0|^3}\;,
\end{equation}
to leading order in the far field. Note that the contributions for the image object follow the same scaling.

Let us now turn to the region $\Omega_3$. There, an apparent singularity at ${\bf r}={\bf r'}$ requires some care. We thus integrate by parts Eq.~\eqref{eq:app:rho3bis} to get:
\begin{align}
        -\frac{\mu}{D_{\rm eff}}\int_{\partial \Omega_3} d^2 {\bf r'} \partial_j' G_N(x,y;x',y') \delta\Sigma_{ij}'\\ + \frac{\mu}{D_{\rm eff}}\int_{\Omega_3} d^2 {\bf r'} \partial_j' G_N(x,y;x',y') \partial_i'\delta\Sigma_{ij}'\label{eq:app:rho3ter}\;.
\end{align}
Let us first look at the boundary term. On $\partial\Omega_3$, $|{\bfr'-\bfr_0}|>|{\bfr-\bfr_0}|/2$ so that $|\delta\Sigma_{ij}'|$ decays asymptotically as $\mathcal{O}(|{\bfr-\bfr_0}|^{-3})$. Furthermore, the gradient of the Green's function scales as $|\partial_i' G_N|\sim R_3^{-1}$. Overall, the boundary integral can thus be controlled as:
\begin{equation*}
    \int_{\partial \Omega_3} d^2 {\bf r'} \partial_j' G_N(x,y;x',y') \delta\Sigma_{ij}'={\cal O}\left( \frac{P(\Omega_3)}{R_3} \frac{1}{|\bfr-\bfr_0|^3} \right) \;,
\end{equation*}
where $P(\Omega_3)=2 \pi R_3$ is the length of $\partial\Omega_3$. This integral is thus negligible compared to the contribution of Eq.~\eqref{eq:app:rho2}.

To estimate Eq.~\eqref{eq:app:rho3ter}, we note that it can estimated as
\begin{equation}\label{eq:Sigma contribution}
    \intop_{\Omega_3} d^2\bfr'\, \left[\partial_j'G_N(\bfr;\bfr')\right] \left[\partial_i'\delta\Sigma_{ij}'\right] \sim \intop_{\Omega_3} \frac{d^2\bfr'}{|\bfr-\bfr'|} \frac{1}{|\bfr'-\bfr_0|^4}\;.
\end{equation}
To show that there is no divergence at $\bfr=\bfr'$ we change variable to $\vec{u}\equiv\bfr-\bfr'$. With this, Eq.~\eqref{eq:Sigma contribution} becomes
\begin{equation}\label{eq:integral u}
    \intop_{\Omega_3} \frac{u du d\gamma}{u} \frac{1}{|\bfr-\bfr_0-\vec{u}|^4}\;,
\end{equation}
with $u=|\vec{u}|$ and $\tan\gamma=u_y/u_x$, showing that the $u\to0$ limit is regular. The $|\bfr-\bfr_0|^4$ scaling then shows the contribution of Eq.~\eqref{eq:Sigma contribution} to be negligible compared to Eq.~\eqref{eq:app:rho2}.

We are finally left with the integral over region $\Omega_2$, which scales as
\begin{equation}\label{eq:Sigma contribution2}
    \intop_{\Omega_2} d^2\bfr'\, \left[\partial_i' \partial_j'G_N(\bfr;\bfr')\right] \delta\Sigma_{ij} \sim \intop_{\Omega_2} \frac{d^2\bfr'}{|\bfr-\bfr'|^2} \frac{1}{|\bfr'-\bfr_0|^3}\;.
\end{equation}
Since, in this region, $|\bfr-\bfr'|>R_3$, this integral is smaller than:
\begin{equation}\label{eq:Sigma contribution3}
    \frac{1}{R_3^2}\intop_{\Omega_2} \frac{d^2\bfr'}{|\bfr'-\bfr_0|^3}\;.
\end{equation}
The last integral can then be extended over the complementary to $\Omega_1$ in $\mathbb{R}^2$, leading to a contribution smaller than
\begin{equation}\label{eq:Sigma contribution4}
    \left|\,\intop_{\Omega_2} d^2\bfr'\, \left[\partial_i' \partial_j'G_N(\bfr;\bfr')\right] \delta\Sigma_{ij}\right| \lesssim \frac{1}{R_1 R_3^2}\;.
\end{equation}
Since $R_3=\varepsilon |\bfr-\bfr_0|$, this final contribution is also negligible compared to Eq.~\eqref{eq:app:rho2}. This concludes our demonstration that Eq.~\eqref{eq:app:rho2} is, self-consistently, the dominating contribution to the density modulation.

\bibliography{LocalizationBib.bib}

\end{document}